\documentclass[preprint,nofootinbib,aps,superscriptaddress,eqsecnum]{revtex4-1}
 \pdfoutput=1
\usepackage{cancel}

\usepackage{float}
\usepackage{graphicx}
\usepackage{amsmath} 
\usepackage{amsfonts}
\usepackage{amssymb}
\usepackage[bookmarks=true,bookmarksopen=true,
                    bookmarksnumbered=true,
                    colorlinks=true,linkcolor=black,
                    citecolor=red]{hyperref}
\usepackage{subfigure}
\usepackage{slashed}
 \usepackage{setspace} 
\usepackage{caption}
\usepackage[normalem]{ulem}
\usepackage{xcolor}
\allowdisplaybreaks
\def\bea{\begin{eqnarray}}
\def\eea{\end{eqnarray}}
\def\be{\begin{equation}}
\def\ee{\end{equation}}
\def\nn{\nonumber}

\newcommand{\ifb}{{\rm fb}^{-1}}

\begin{document}
\title{\LARGE{Prospects for detecting the rare heavy Higgs decay $H\to h\gamma\gamma$ through the $H\to b\bar{b}\gamma\gamma$ channel at the LHC}}

\author{M. A. Arroyo-Ure\~na}
\email{marco.arroyo@fcfm.buap.mx}
\affiliation{Facultad de Ciencias F\'isico-Matem\'aticas, \\
Benem\'erita Universidad Aut\'onoma de Puebla,\\
 C.P. 72570, Puebla, Pue., M\'exico.}
 \affiliation{Centro Interdisciplinario de Investigaci\'on y Ense\~nanza de la Ciencia,\\
Benem\'erita Universidad Aut\'onoma de Puebla,\\
 C.P. 72570, Puebla, Pue., M\'exico.}
 
\author{Alejandro Ibarra}
\email{ibarra@tum.de}
\affiliation{Physik-Department, Technische Universität München, James-Franck-Straße, 85748, Garching, Germany.}

\author{Pablo Roig}
\email{pablo.roig@cinvestav.mx, paroig@ific.uv.es}
\affiliation{Departamento de F\'\i sica, Centro de Investigaci\' on y de Estudios Avanzados del Instituto Polit\' ecnico Nacional, Apartado Postal 14-740, 07000 Ciudad de M\'exico, M\'exico}
\affiliation{IFIC, Universitat de Val\`encia – CSIC, Catedr\'atico Jos\'e
Beltr\'an 2, E-46980 Paterna, Spain.}

\author{T. Valencia-P\'erez}
\email{tvalencia@fisica.unam.mx}
\affiliation{Instituto de F\'isica, Universidad Nacional Aut\'onoma de M\'exico, 01000, CDMX, México.}
 \begin{abstract}
 	We study the decay of a heavy CP-even neutral Higgs into an on-shell Standard Model-like Higgs boson and two photons, $H\to h\gamma\gamma$, in the two-Higgs doublet model. We argue that the decay channel $H\to h\gamma\gamma$, followed by the decay of the Standard Model Higgs $h\rightarrow b\bar b$, could be observed at the 5$\sigma$ level at the High-Luminosity LHC for masses of the heavy Higgs up to  950 GeV for the type-II, 650 GeV for the Lepton Specific and the Flipped 2HDMs, and 350 GeV  for the type-I. We also discuss the possible role of the decay $H\to h\gamma\gamma$ in discriminating among different types of 2HDMs and in enhancing the total number of events in the final state $H\rightarrow b\bar b \gamma\gamma$ compared to the cascade decay $H\to hh$ followed by $h\to\gamma\gamma$ $h\to b\bar{b}$ with identical final state (although with different kinematical distributions).
\end{abstract}

\maketitle

\section{Introduction}

A bit more than a decade ago, the ATLAS and CMS collaborations announced the observation of a new spin-0 particle, with properties consistent with those expected for the Higgs boson predicted in the mid-1960s by Brout, Englert, and Higgs, as a remnant of the spontaneous breaking of the electroweak symmetry by the vacuum expectation value of a complex scalar, doublet under $SU(2)_L$ and with opposite hypercharge to the lepton doublet \cite{ATLAS:2012yve, CMS:2012qbp}. The observation of the Higgs boson has stimulated the search for heavier copies of the Higgs field (akin to the heavier generations existing in the fermion sector), or other fundamental scalar particles in other representations. 

One of the simplest extensions of the Standard Model Higgs sector is the Two Higgs Doublet Model (2HDM), which predicts the existence of one CP even neutral ($H$), one CP odd neutral ($A$) and two charged Higgs scalars ($H^\pm$)\cite{Lee:1973iz,Georgi:1978xz,Deshpande:1977rw,Barnett:1983mm}. These particles carry electroweak charges and therefore could be produced in collider experiments and detected through their decay into Standard Model particles. The ATLAS and CMS collaborations performed an intensive search for heavy neutral and charged Higgs bosons $(H,\,A,\,H^{\pm})$ in a broad range of masses for the new particles and for different decay channels (see e.g. refs.~\cite{ATLAS:2021ifb, ATLAS:2023zkt
,CMS:2016cma,CMS:2023boe}~\footnote{See also the PDG information on the searches for neutral and charged Higgs bosons \cite{ParticleDataGroup:2024cfk} and the ATLAS and CMS webpages, \url{https://atlas.web.cern.ch/Atlas/GROUPS/PHYSICS/PUBNOTES/ATL-PHYS-PUB-2024-008/} and \url{https://cms-results.web.cern.ch/cms-results/public-results/publications/HIG/SUS.html}, respectively.}). The Gfitter group updated their global electroweak fit and examined the constraints on 2HDMs in ref.~\cite{Haller:2018nnx}. This work concludes that the compatibility with the measured values of the Higgs coupling, along other electroweak observables, requires a strong degeneracy of either the $H$ or the $A$ boson mass with the $H^\pm$ boson mass (see also e.g. refs.~\cite{Wang:2022yhm}). 

In this work we focus on ``generation" changing processes in the Higgs sector involving photons in the final state. We are motivated by analogous processes  in the fermionic sector, such as $b\rightarrow s\gamma$ and $\mu\rightarrow e\gamma$, which provide clean signatures and which are commonly used as tests of the Standard Model. For spin-0 particles the decay $H\rightarrow h \gamma$ is forbidden by Lorentz invariance, therefore we will focus on the three body process  $H \to h \gamma\gamma$ (see also \cite{Ghosh:2019jzu} for a related analysis in the dark sector). Specifically, we will study it with the subsequent decay $h\to b\bar{b}$. To our knowledge, the interest of the decay $H\to b\bar{b}\gamma\gamma$ has not been put forward before searching for a second Higgs doublet. In this context, we will highlight its potential role in discriminating the 2HDM type, complementarily to other observables.

This work is organized as follows. In section~\ref{Sec:Model} we briefly review the 2HDM and the current constraints on its parameter space. Then, in section \ref{Sec:WidthDecay} we describe the calculation of the decay width of the process $H\to h\gamma\gamma$, and in section \ref{Sec:Collider} we discuss the prospects of detecting this decay at the LHC and at the High Luminosity LHC via the $b\bar{b}\gamma\gamma$ final state. Finally, in Section \ref{Sec:conclusions} we present our conclusions. We also include appendices \ref{Acoplos},  \ref{ObParam} and \ref{amplitudes} with the Feynman rules for the 2HDM relevant to calculate the rate for $H\to h\gamma\gamma$, the expressions for the oblique parameters, and the exact one-loop expression for the modulus squared of the amplitude (summing over the final photon polarizations). 


\section{The Two-Higgs Doublet Model}\label{Sec:Model}

Two-Higgs Doublet Models are among the simplest extensions of the SM and have been intensively studied in the literature, {\it e.g.} \cite{Branco:2011iw,Gunion:1989we,Craig:2013hca,Gunion:2002zf}. In these models, the scalar sector consists of two complex $SU(2)$ Higgs doublets $\Phi_a^T=(\phi_a^+,\,\phi_a^0)$ $(a=1,\,2)$ with hypercharge $+1$. The part of the Lagrangian involving the Higgs doublets reads:
\begin{equation}
\mathcal{L}^{\text{2HDM}}_{\Phi_1,\Phi_2}=\sum_a|D_\mu\Phi_a|^2-V(\Phi_1,\,\Phi_2)+\mathcal{L}_{\text{Yuk}},
\end{equation}  
where $D_\mu$ is the SM covariant derivative, $V(\Phi_1,\,\Phi_2)$ is the scalar potential, and $\mathcal{L}_{\text{Yuk}}$ describes the Yukawa interactions of the Higgs doublets to the SM fermions. The scalar potential reads:
\begin{align}\label{potential}
V&=m_{11}^2\Phi_1^{\dagger}\Phi_1+m_{22}^2\Phi^{\dagger}_2\Phi_2-(m_{12}^2\Phi_1^{\dagger}\Phi_2+\rm{H.c.})\nonumber\\
&+\frac{\lambda_1}{2}(\Phi_1^{\dagger}\Phi_1)^2+\frac{\lambda_2}{2}(\Phi_2^{\dagger}\Phi_2)^2+\lambda_3(\Phi_1^{\dagger}\Phi_1)(\Phi_2^{\dagger}\Phi_2)+\lambda_4(\Phi_1^{\dagger}\Phi_2)(\Phi_2^{\dagger}\Phi_1)\nonumber\\
&+\Bigg[\frac{\lambda_5}{2}(\Phi_1^{\dagger}\Phi_2)^2+\lambda_6(\Phi_1^{\dagger}\Phi_1)(\Phi_1^{\dagger}\Phi_2)+\lambda_7(\Phi_2^{\dagger}\Phi_2)(\Phi_1^{\dagger}\Phi_2)+\rm{H.c.}\Bigg].
\end{align}
The parameters $m_{11}^2,\,m_{22}^2$ and $\lambda_{1,\,2,\,3,\,4}$ are real; while, in general, $m_{12}^2$ and $\lambda_{5,\,6,\,7}$ are complex. In what follows we will impose for simplicity the conservation of CP in the scalar sector, so that all parameters are real. Besides, the most general Yukawa Lagrangian for 2HDM is \cite{Branco:2011iw}
\begin{equation}\label{YukLagrangian}
-\mathcal{L}_{\text{Yuk}}=\sum_{a=1}^2\Bigg[\bar{L}_L\Phi_a Y_a^{\ell}\ell_R+\bar{Q}_L\Phi_a Y_a^{d}D_R+\bar{Q}_L\tilde{\Phi}_a Y_a^{u}U_R  \Bigg]+\text{H.c.},
\end{equation}
where $Y_a^{f=\ell,D,U}$ are $3\times 3$ complex matrices, the left-handed fermion doublets ($Q,L$) as well as the right-handed singlets ($\ell,D,U$) are 3-vectors in flavor space, and  $\tilde{\Phi}_a=i\sigma_2\Phi_a^*$.

In order to prevent tree-level flavor changing neutral currents, it is common to impose an exact $Z_2$ symmetry in the Higgs doublets and the right-handed fermion fields. By convention, one adopts $\Phi_2\rightarrow \Phi_2$, $\Phi_1\rightarrow -\Phi_1$ and $U_R\rightarrow U_R$, so that the right-handed quarks with charge $2/3$ couple only to $\Phi_2$. On the other hand, the right-handed quarks with charge $-1/3$ and the right-handed leptons can be even or odd under the $Z_2$ symmetry, leading to  only four possible discrete symmetries suppressing the tree-level flavor-changing neutral currents (FCNC)
\cite{Barger:2009me,Barnett:1983mm}.
These are summarized in Table \ref{2HDMtypes}, and are named type I (2HDM-I), type II (2HDM-II), lepton specific (2HDM-LS) and flipped (2HDM-F). Note that the $Z_2$ symmetry forces the terms proportional to $\lambda_{6,\,7}$ and $m_{12}$ to vanish in Eq. \eqref{potential}. On the other hand, we will allow in our analysis the soft breaking of the $Z_2$ symmetry in the potential, which allows the term proportional to $m_{12}$.

When the neutral components of the Higgs fields get vacuum expectation values (VEVs), the electroweak symmetry breaks spontaneously. The two scalar doublets can be expressed in terms of real fields and the VEVs $v_1$ and $v_2$
\begin{align}
\Phi_{a}=\begin{pmatrix}
\phi_a^{+}\\
\displaystyle{\frac{v_{a}+
\rho_{a}+i\eta_{a}}{\sqrt{2}}}
\end{pmatrix},
\end{align}
where it is common to relate $v_a$ to the VEV of the Higgs field in the Standard Model, $v=246$ GeV as $v_1= v\cos\beta$ and $v_2=v \sin\beta$.

Three of the eight degrees of freedom in the Higgs doublets are absorbed as longitudinal components of the $W^{\pm}$ and $Z^0$ gauge bosons, while the remaining five degrees of freedom include three electrically neutral scalars and a pair of charged scalars. Upon diagonalizing the mass matrices, these five degrees of freedom are recast as the SM-like Higgs boson $h$ and a heavy scalar $H$ (with masses $m_h<M_H$), one CP-odd scalar $A$ (with mass $M_A$) and a pair of charged scalar bosons $H^{\pm}$ (with mass $m_{H^\pm}$). The mass eigenstates of the neutral particles of the model are related to the interaction eigestates by $h=\rho_1\sin \alpha -\rho_2\cos\alpha$, $H=-\rho_1\cos\alpha-\rho_2\sin\alpha$, $A=\eta_1 \sin\beta-\eta_2 \cos\beta$.

Replacing the interaction eigenstates by mass eigenstates in 
Eq.~(\ref{YukLagrangian}) one finds 
\begin{eqnarray}\label{Lcharged}
-\mathcal{L}_{\text{Yukawa}}^{\text{2HDM}}&=&\sum_{f=U,\,D,\,\ell}\frac{m_f}{v}\Bigg(g_h^f\bar{f}fh+g_H^f\bar{f}fH -i g_A^f\bar{f}fA\gamma_5\Bigg)\nonumber\\
&+&\Bigg[\frac{\sqrt{2}V_{ij}}{v}\bar{U}_i(m_U g_{H^+}^uP_L+m_D g_{H^+}^dP_R)D_jH^{+} + \frac{\sqrt{2}m_{\ell}g_{H^+}^{\ell}}{v}\bar{\nu}_{\ell\,L} \ell_RH^+ +\text{H.c.}\Bigg],
\end{eqnarray}
where $P_{L\,(R)}$ is the projection operator for left- (right-) handed fermions ($U=u,c,t;D=d,s,b;\ell=e,\mu,\tau$), and the coupling constants $g_{\phi}^f$ ($\phi=A,\,H,\,h,\,H^{\pm}$) are given in Table \ref{Couplings}. 
Further, rotating similarly the kinetic terms and the scalar potential, one finds interaction terms of the various scalar mass eigenstates with the electroweak gauge bosons, as well as interactions among the different scalar fields from the 2HDM potential. The Feynman rules for these interaction terms are compiled in Appendix \ref{Acoplos}. It is important to note that the couplings of the spin-0 particles to the fermions depend on the type of 2HDM, whereas the couplings among the bosons (spin-0 and/or spin-1) are universal for all 2HDMs.

In phenomenological analyses it is common to replace the eight free parameters of the potential $m^2_{11}$, $m^2_{22}$,  $m^2_{12}$ and $\lambda_i$, $i=1...5$, by the following set of parameters: $v$, $m_h$, $M_H$, $M_A$, $M_{H^\pm}$, $\tan\beta$, $\cos(\beta-\alpha)$ and $m_{12}^2$ (for explicit expressions, see appendix D of ref.~\cite{Gunion:2002zf}). These parameters are subject to a number of experimental and theoretical constraints, which include:
\subsection*{Experimental constraints}
\begin{itemize}
\item Resonant production of neutral scalars in proton-proton collisions.\\
The SM-like Higgs boson, $h$, can be resonantly produced at the LHC, $pp\rightarrow h$, and then decay into Standard Model particles. It is parameterized through the ratio 
\begin{equation}
	\mathcal{R}_{X}=\frac{\sigma(pp\to h)\cdot\mathcal{BR}(h\to X)}{\sigma(pp\to h^{\text{SM}})\cdot\mathcal{BR}(h^{\text{SM}}\to X)},
\end{equation}
here $h$ is the SM-like Higgs boson coming from an extension of the SM.
The production cross-section and the decay rates $\sigma(pp\to h)$ and $\mathcal{BR}(h\to X)$, respectively, depend on the angles $\alpha$ and $\beta$. Given the excellent agreement of the Standard Model with the experimental results on the production of $X=b\bar{b},\;\tau^-\tau^+,\;\mu^-\mu^+,\;WW^*,\;ZZ^*,\;\gamma\gamma$ pairs by a 125 GeV mass resonance, it is possible to set constraints on $\tan\beta$ and $\cos(\beta-\alpha)$~\cite{ATLAS:2019nkf}.

Similarly, the heavy CP-even scalar $H$ can contribute to the production of a pair of Higgs bosons, which subsequently decay into $b\bar b \gamma\gamma$,  {\it i.e.}, $\sigma(pp\to H\to hh\to b\bar{b}\gamma\gamma)$~\cite{ATLAS:2021ifb}. This observable can bind $M_H$ and at the same time discern the production process of the final state $\bar{b}b\gamma\gamma$ that arises from a pure two-body process $(hh)$ and the genuine three-body decay $h\gamma\gamma$. Figure~\ref{XS_H-hh} shows the $\sigma(pp\to H\to hh\to \bar{b}b\gamma\gamma)$ as a function of $M_H$. We notice that the versions 2HDM-I, LS and Flipped are favored in the region of small $\tan\beta\in\left[0.1,1\right]$ while the 2HDM-II motivates the study of larger $\tan\beta\in\left[1,5\right]$. It is important to note that $\tan\beta=0.1$ excludes $M_H\in\left[300,900\right]$, $M_H\in\left[300,600\right]$, $M_H\in\left[300,600\right]$ for 2HDM-I, LS and Flipped, respectively. For 2HDM-II, $\tan\beta=8$ rules outs $900<M_H$ GeV. We implement the relevant interactions via $\texttt{LanHEP}$~\cite{Semenov:2014rea} for $\texttt{MadGraph5}$ \cite{Alwall:2014hca} to evaluate the leading order cross sections at a center-of-mass energy $\sqrt{s}=13$ TeV by considering the $\texttt{NN23LO1}$ PDF set \cite{NNPDF:2017mvq}.
	 It is worth mentioning that, for typical benchmark points, $\sigma(pp\to H\to hh\to b\bar{b}\gamma\gamma)$ is 
    a factor $\sim 4$ larger than $\sigma(pp\to H\to h\gamma\gamma,\,h \to b\bar{b})$, as will be seen later (cf. figs.~\ref{XS_H-hh} and \ref{XSallModels}).
\begin{figure}[!htb]
	\centering
	\subfigure[]{{\includegraphics[scale=0.1865,angle=0]{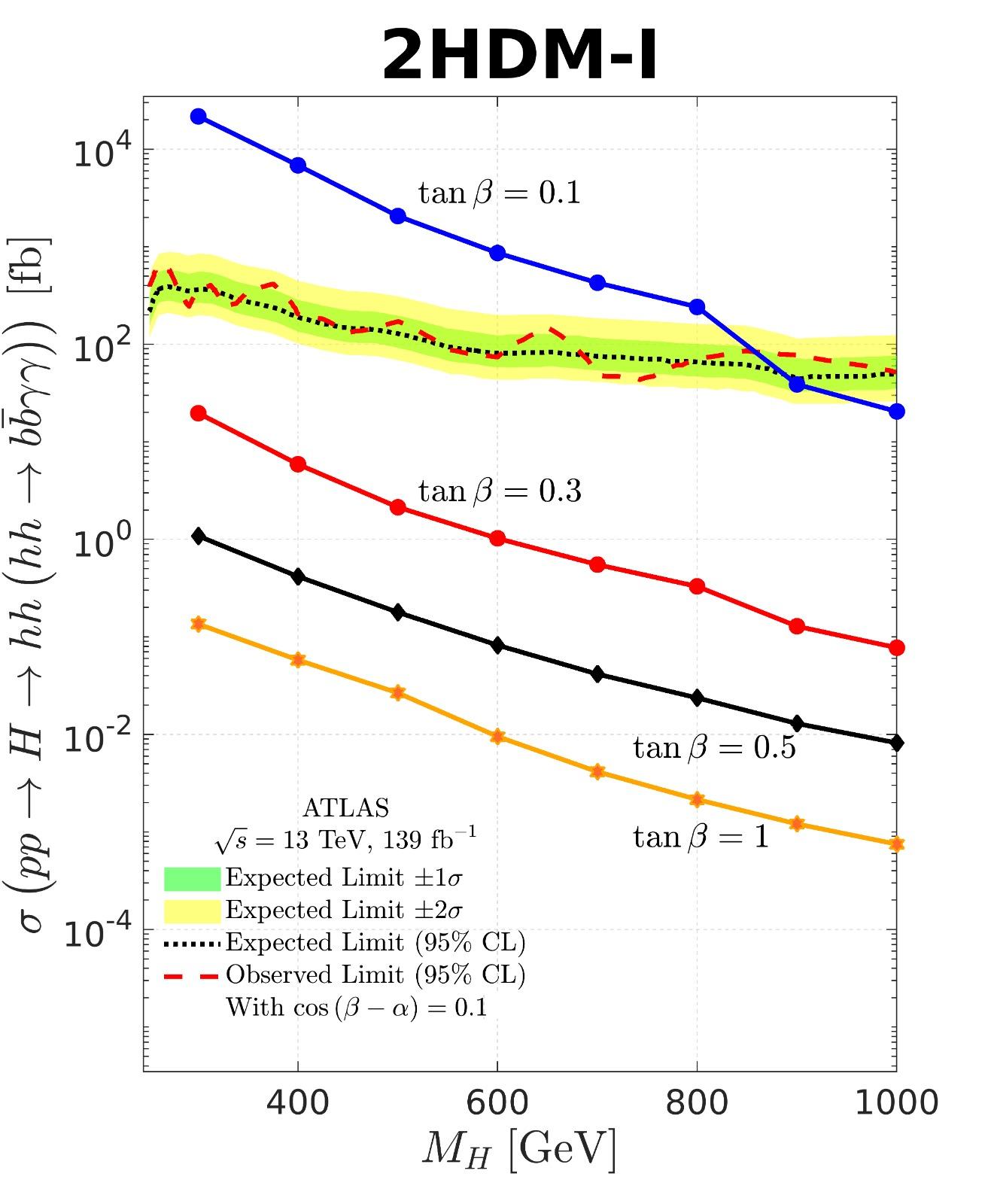}}}
	\subfigure[]{{\includegraphics[scale=0.14,angle=0]{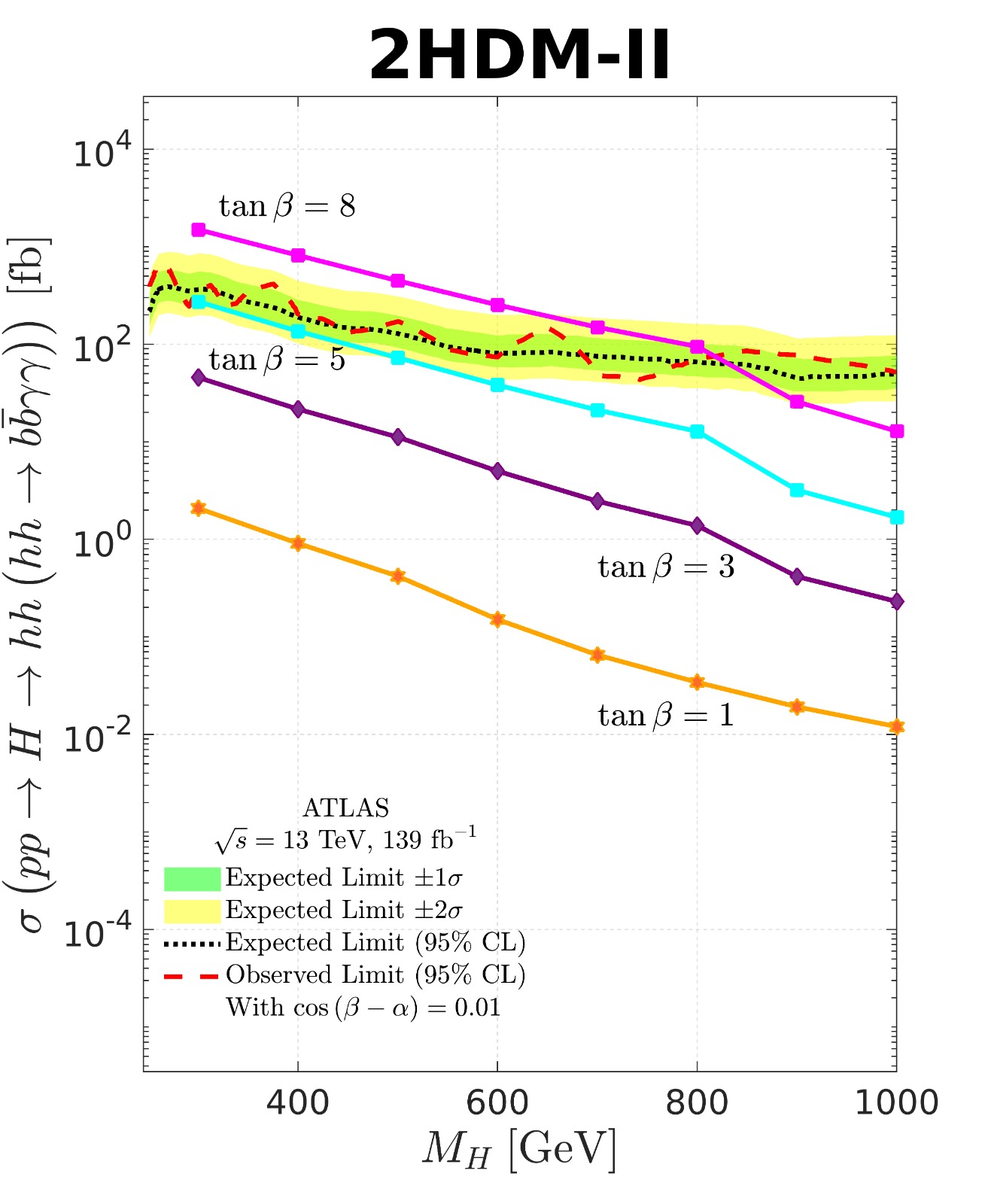}}}
	\subfigure[]{{\includegraphics[scale=0.14,angle=0]{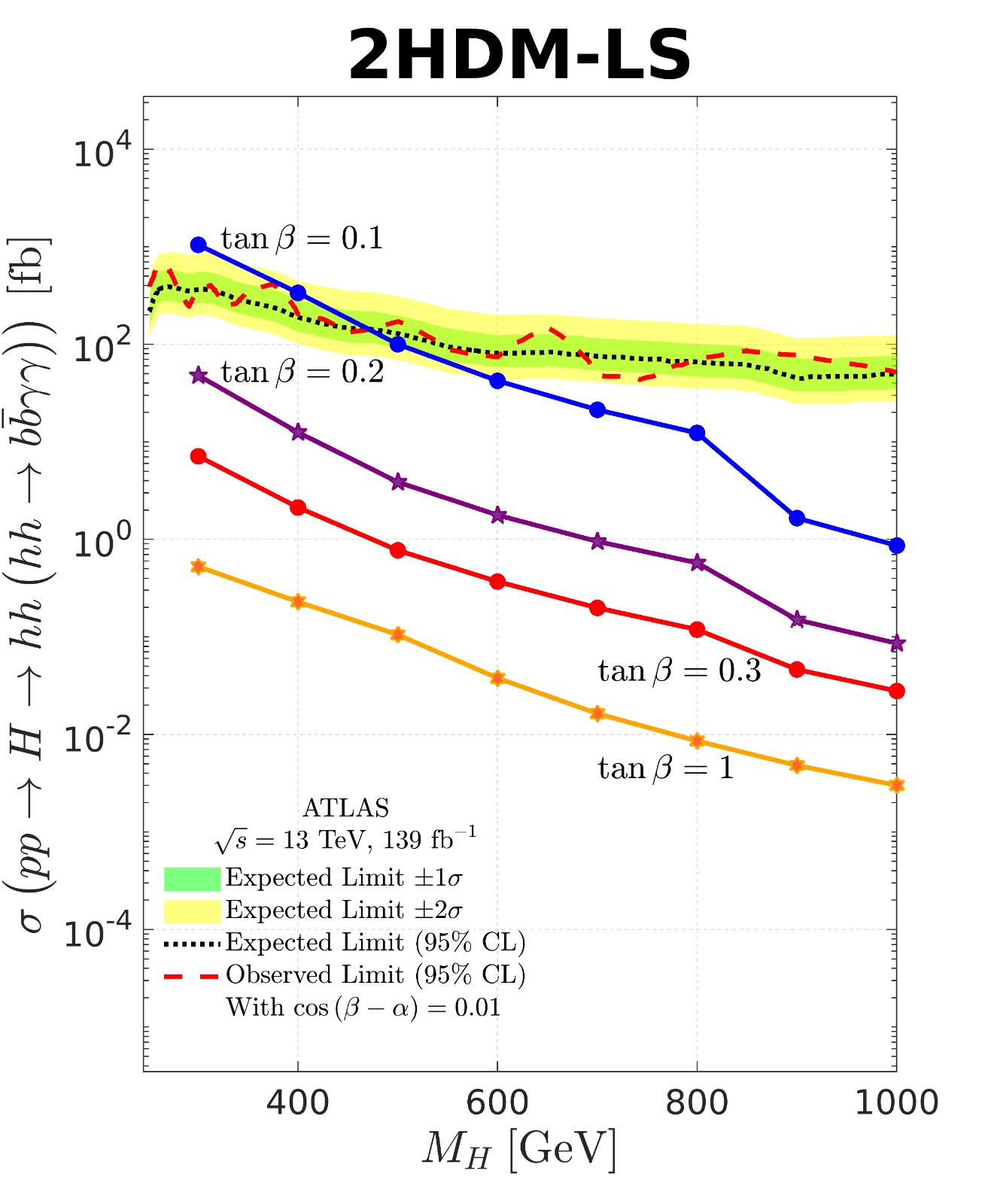}}}
	\subfigure[]{{\includegraphics[scale=0.14,angle=0]{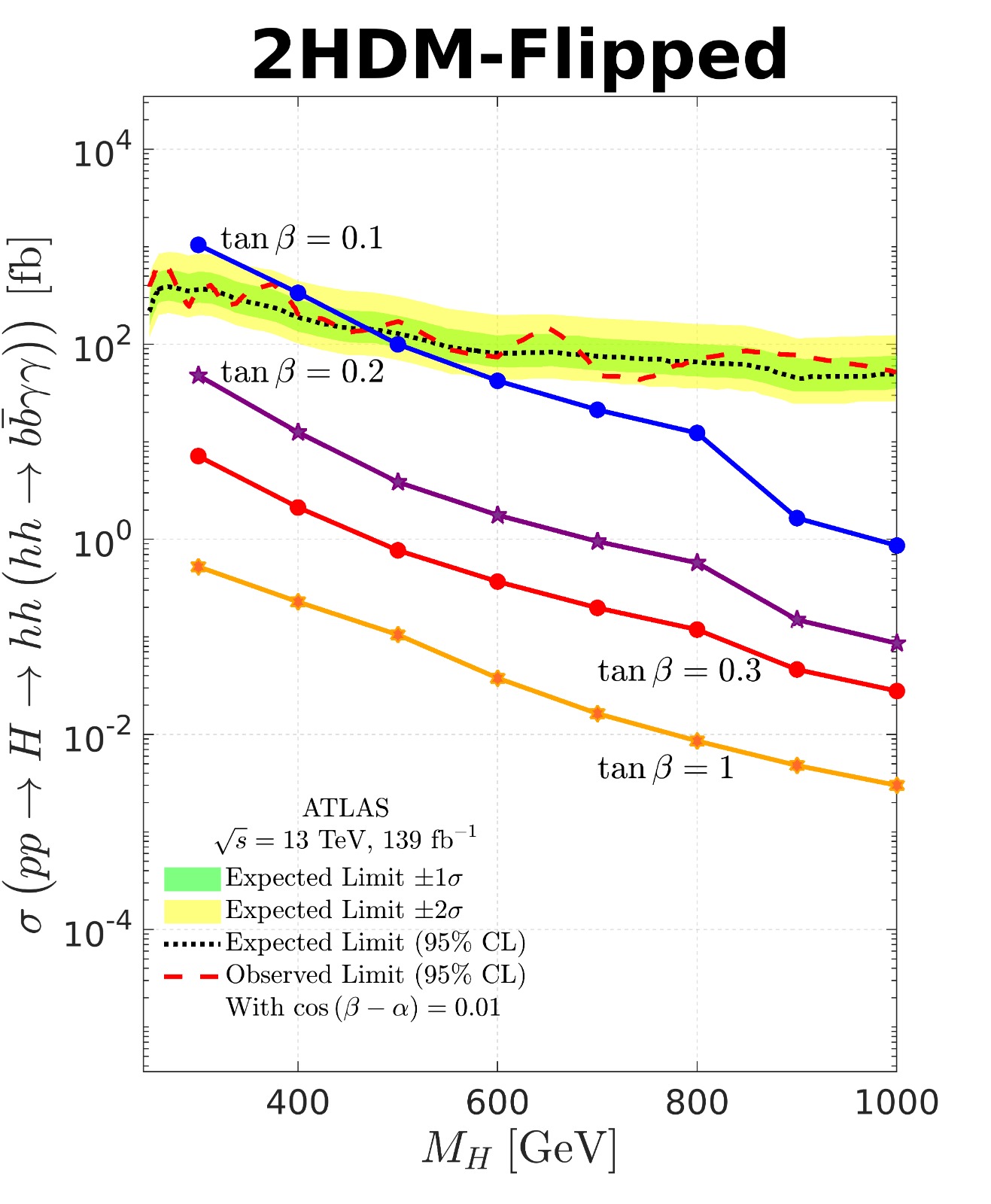}}}
	\caption{Observed and expected limits at 95\% CL on the production cross section of a narrow-width scalar
		resonance $H$ as a function of the mass $M_H$ of the hypothetical scalar particle. The black solid line represents the observed upper limits. The dashed line represents the expected upper limits. The $\pm 1\sigma$ and $\pm 2\sigma$ variations about the
		expected limit due to statistical and systematic uncertainties are also shown. (a) 2HDM-I, (b) 2HDM-II, (c) 2HDM Lepton Specific, (d) 2HDM-Flipped.}\label{XS_H-hh}
\end{figure}

\item $b\rightarrow s\gamma$\\
The 2HDMs considered in this paper do not induce flavor changing neutral currents at tree level; however, processes changing flavor can be induced at the quantum level through the exchange of charged scalars in loops. In our analysis, we consider the rare decay $b\rightarrow s\gamma$  \cite{Ciuchini:1997xe, Misiak:2017bgg} through the ratio
\begin{equation}\label{Rquark}
	R_{\rm quark}=\frac{\Gamma(b\to X_s \gamma)}{\Gamma(b\to X_c e\nu_e)}=(3.22\pm0.15)\times 10^{-3},
\end{equation}
with $\mathcal{BR}(B^+\to X_c e^+\nu_e)=(10.8\pm 0.4)\%$ \cite{
ParticleDataGroup:2024cfk}.
We have implemented in a \texttt{Mathematica}\footnote{We make our code available upon request.} code the analytical expressions reported in \cite{Ciuchini:1997xe} for reproducing the ratio in Eq.~\eqref{Rquark} and we made sure that the scanned masses $M_{H^{\pm}}$, as shown in Table~\ref{scan}, are actually allowed by $R_{\rm quark}$.
\subsection*{Theoretical constraints}

\item Perturbativity \\
We impose the requirement that no scalar coupling exceeds $4\pi$, so that our perturbative analysis remains valid. This imposes
\begin{align}\label{Perturbativity}
		{\rm Max}\Bigg\{|\lambda_{1,2,3,4,5}|,|\lambda_3+\lambda_4\pm\lambda_5|, |\lambda_4\pm \lambda_5|/2, |\lambda_3+\lambda_4|\Bigg\}\leq 4\pi\,.
\end{align}
\item Vacuum stability\\
We require that the potential, Eq. \eqref{potential}, is bounded from below and stays positive for arbitrarily large values of the fields. This requirement translates into:
	\begin{equation}\label{VS}
		\lambda_1>0, \lambda_2>0, \lambda_3+\sqrt{\lambda_1\lambda_2}>0, \lambda_3+\lambda_4-|\lambda_5|+\sqrt{\lambda_1\lambda_2}>0, 
	\end{equation}
 where the parameters $\lambda_i$ ($i=1,\,2,...,5$) are given in Eq. \eqref{potential}.
\item Unitarity of the $S$-matrix.\\
We demand that the $S$-matrix on scalar to scalar, gauge boson to gauge boson, and scalar to gauge boson scatterings remain unitary at the perturbative level~\cite{Goodsell:2018fex}
	\begin{eqnarray}\label{Unitarity}
		\text{Max}&&\Bigg\{|\lambda_3\pm\lambda_4|, |\lambda_1+\lambda_2\pm\sqrt{(\lambda_1-\lambda_2)^2+\lambda_4^2}|, |\lambda_3\pm\lambda_5|, \nonumber\\ 
		&&|3(\lambda_1+\lambda_2)\pm\sqrt{9(\lambda_1-\lambda_2)^2+(2\lambda_3+\lambda_4)^2}|\\
		&&|\lambda_3+2\lambda_4\pm3\lambda_5|, |\lambda_1+\lambda_2\pm\sqrt{(\lambda_1-\lambda_2)^2+\lambda_5^2}|
		\Bigg\}<8\pi.\nonumber
\end{eqnarray}
\item Oblique parameters\\
We require that the Peskin-Takeuchi parameters S, T are within the experimental measurements, $S=-0.01\pm 0.07$, $T=0.04\pm 0.06$ (assuming $U=0$, which is approximately satisfied in the 2HDM). The expressions of the three parameters S, T, and U are collected in Appendix \ref{ObParam}.
\end{itemize}

\begin{figure}[!h]
	\centering
	\subfigure[]{{\includegraphics[scale=0.25]{
 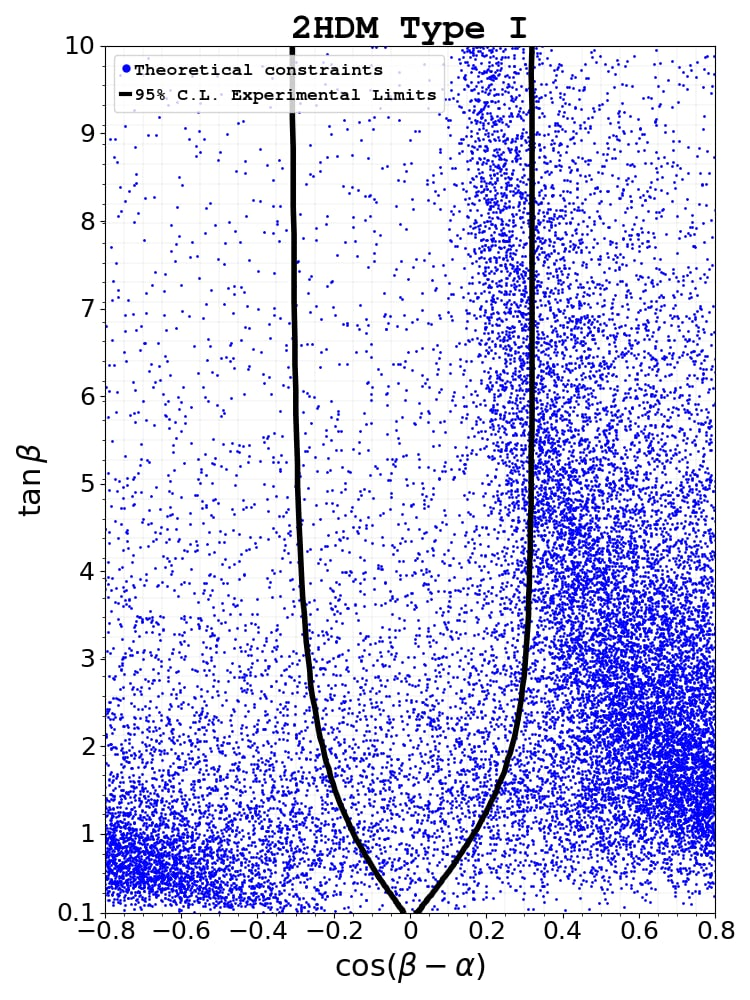}}}\label{a}
	\subfigure[]{{\includegraphics[scale=0.25]{
 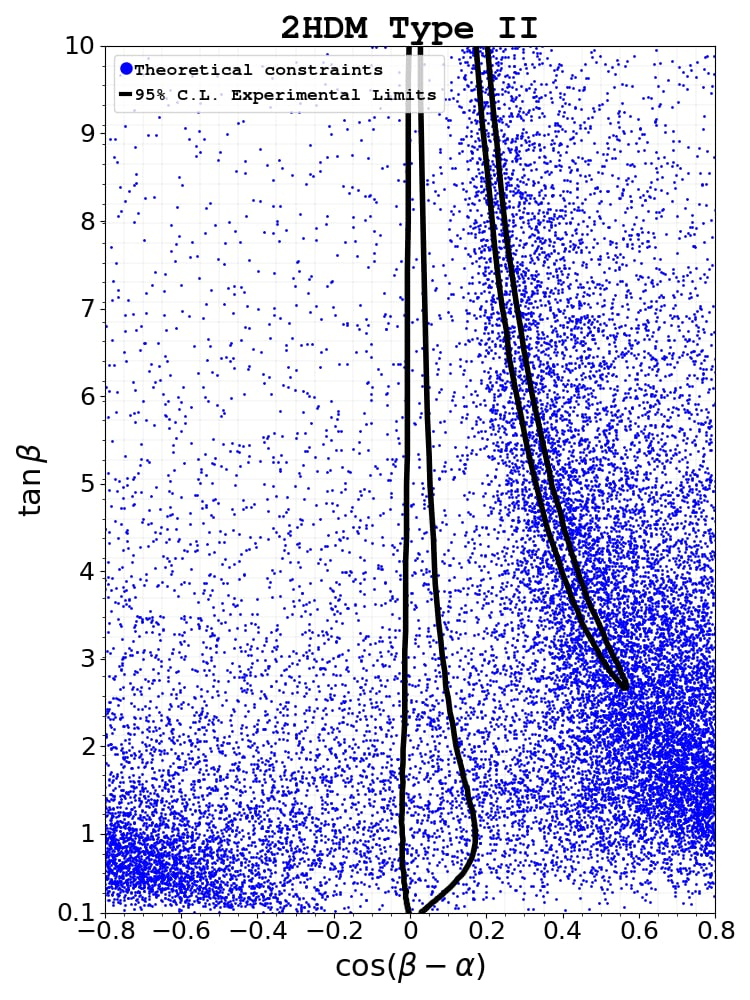}}}\label{b}
	\subfigure[]{{\includegraphics[scale=0.25]{
 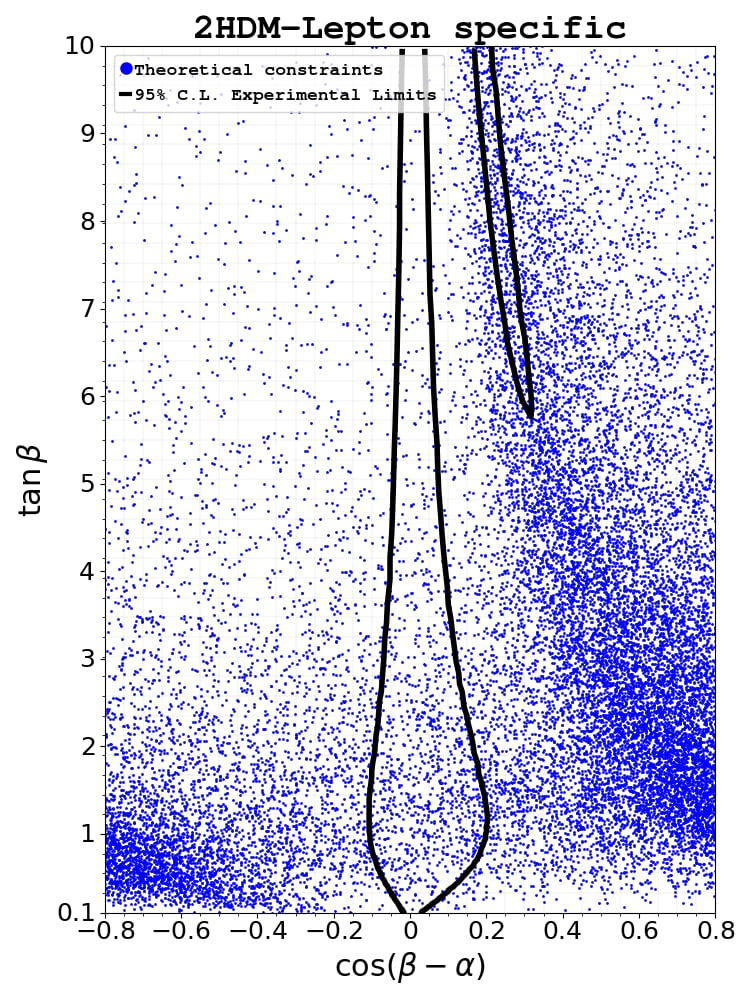}}}\label{c}
	\subfigure[]{{\includegraphics[scale=0.25]{
 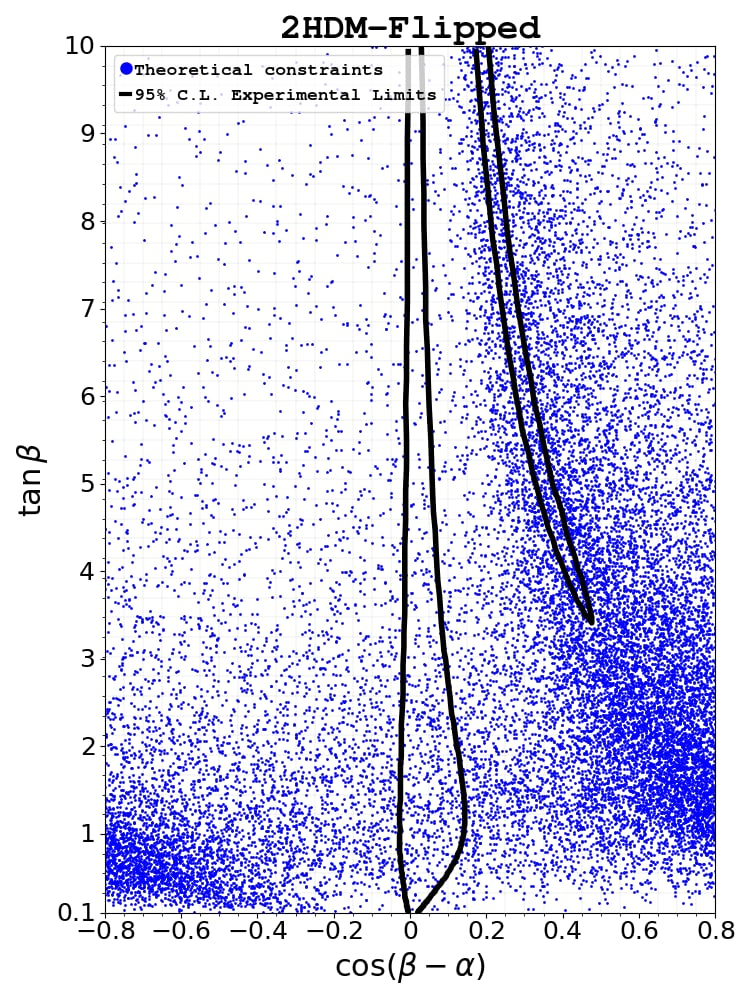}}}\label{d}
	\caption{Allowed $\cos(\beta-\alpha)-\tan\beta$ plane for different versions of 2HDM, assuming $m_{12}=M_{A}=M_{H^{\pm}}=M_{H}=700$ GeV: (a) Type I, (b) Type II, (c) Lepton Specific, (d) Flipped.}\label{AllConstraints}
\end{figure}

We present in Fig. \ref{AllConstraints} the regions in the parameter space spanned by $\cos(\beta-\alpha)$ and $\tan\beta$ that satisfy the above experimental and theoretical constraints in the four 2HDMs. The area limited by the black curves indicates the regions allowed by the LHC and by flavor physics at 95\% C.L. which were generated via the \texttt{Mathematica} package so-called \texttt{SpaceMath}~\cite{Arroyo-Urena:2020qup}. Further, points in the parameter space that satisfy the theoretical constraints are displayed in blue. We present in Table~\ref{scan} the range of scanned values for the free parameters such that they satisfy all theoretical constraints; the remainder of the parameter space is mostly excluded by $\lambda_3+\lambda_4-|\lambda_5|+\sqrt{\lambda_1\lambda_2}>0$, $|3(\lambda_1+\lambda_2)\pm\sqrt{9(\lambda_1-\lambda_2)^2+(2\lambda_3+\lambda_4)^2}|<8\pi$ and the oblique parameter $T$. 
\begin{table}[!htb]
	\caption{Interval of parameters scanned.}
		\begin{centering}\label{scan}
		\begin{tabular}{cc}
			\hline 
			Parameter & Scanned range\tabularnewline
			\hline 
			\hline 
			$M_{H,\,A}$ & {[}$m_{h},1000${]} GeV\tabularnewline
			\hline 
			$M_{H^{\pm}}$ & {[}$400,1000${]} GeV\tabularnewline
			\hline 
			$\tan\beta$ & {[}0.001,50{]}\tabularnewline
			\hline 
			$m_{12}$ & {[}100-1000{]} GeV\tabularnewline
			\hline 
			$\cos(\beta-\alpha)$ & {[}-1,1{]}\tabularnewline
			\hline 
		\end{tabular}
		\par\end{centering}
\end{table}

We find that these constraints favor the regions with $\beta\pm\alpha\sim\pi/2$ (see also \cite{Haller:2018nnx}). More concretely, $\beta-\alpha\sim \pi/2$ corresponds to the alignment limit, and the various constraints favor the region $\cos(\beta-\alpha)\simeq 0$ 
and $0.1\lesssim \tan\beta \lesssim 5$. On the other hand,  $\beta+\alpha\sim\pi/2$ leads to a sign flip in the
SM-like Yukawa couplings, yielding a band in the $\tan\beta-\cos(\beta-\alpha)$ plane still allowed by experiments. In our numerical analysis in Section \ref{Sec:Collider}, we will focus on the alignment limit region, which seems physically better motivated. 

\begin{table}
\centering{}%
\begin{tabular}{|cccc|}
\hline 
2HDM & $U_R^i$ & $D_R^i$ & $\ell_R^i$\tabularnewline
\hline 
\hline 
Type-I & $\Phi_{2}$ & $\Phi_{2}$ & $\Phi_{2}$\tabularnewline
\hline 
Type-II & $\Phi_{2}$ & $\Phi_{1}$ & $\Phi_{1}$\tabularnewline
\hline 
Lepton-specific & $\Phi_{2}$ & $\Phi_{2}$ & $\Phi_{1}$\tabularnewline
\hline 
Flipped & $\Phi_{2}$ & $\Phi_{1}$ & $\Phi_{2}$\tabularnewline
\hline 
\end{tabular}
\caption{Two-Higgs Doublet Models which lead to natural flavor conservation (absence of FCNC at tree level). The superscript $i$ indicates the generation index. By convention, the $u_R^i$ always couple to $\Phi_2$.}\label{2HDMtypes}
\end{table}

	\begin{table}
		\begin{centering}
			\begin{tabular}{|c|c|c|c|c|}
				\hline 
				Coupling & Type I & Type II & Lepton-specific & Flipped\tabularnewline
				\hline 
				\hline 
				$g_{h}^{U}$ & $\cos\alpha/\sin\beta$ & $\cos\alpha/\sin\beta$ & $\cos\alpha/\sin\beta$ & $\cos\alpha/\sin\beta$\tabularnewline
				\hline 
				$g_{h}^{D}$ & $\cos\alpha/\sin\beta$ & $-\sin\alpha/\cos\beta$ & $\cos\alpha/\sin\beta$ & $-\sin\alpha/\cos\beta$\tabularnewline
				\hline 
				$g_{h}^{\ell}$ & $\cos\alpha/\sin\beta$ & $-\sin\alpha/\cos\beta$ & $-\sin\alpha/\cos\beta$ & $\cos\alpha/\sin\beta$\tabularnewline
				\hline 
				$g_{H}^{U}$ & $\sin\alpha/\sin\beta$ & $\sin\alpha/\sin\beta$ & $\sin\alpha/\sin\beta$ & $\sin\alpha/\sin\beta$\tabularnewline
				\hline 
				$g_{H}^{D}$ & $\sin\alpha/\sin\beta$ & $\cos\alpha/\cos\beta$ & $\sin\alpha/\sin\beta$ & $\cos\alpha/\cos\beta$\tabularnewline
				\hline 
				$g_{H}^{\ell}$ & $\sin\alpha/\sin\beta$ & $\cos\alpha/\cos\beta$ & $\cos\alpha/\cos\beta$ & $\sin\alpha/\sin\beta$\tabularnewline
				\hline 
				$g_{A}^{U}$ & $\cot\beta$ & $\cot\beta$ & $\cot\beta$ & $\cot\beta$\tabularnewline
				\hline 
				$g_{A}^{D}$ & $-\cot\beta$ & $\tan\beta$ & $-\cot\beta$ & $\tan\beta$\tabularnewline
				\hline 
				$g_{A}^{\ell}$ & $-\cot\beta$ & $\tan\beta$ & $\tan\beta$ & $-\cot\beta$\tabularnewline
				\hline  \hline
				$g_{hVV}$ & $\sin(\beta-\alpha)$ & $\sin(\beta-\alpha)$ & $\sin(\beta-\alpha)$ & $\sin(\beta-\alpha)$\tabularnewline
				\hline 
				$g_{HVV}$ & $\cos(\beta-\alpha)$ & $\cos(\beta-\alpha)$ & $\cos(\beta-\alpha)$ & $\cos(\beta-\alpha)$\tabularnewline
				\hline 
				$g_{AVV}$ & $0$ & $0$ & $0$ & $0$\tabularnewline
				\hline 
				$g_{\gamma_{\alpha}H^{+}H^{-}}$ & $-ie(p'+p)_{\alpha}$ & $-ie(p'+p)_{\alpha}$ & $-ie(p'+p)_{\alpha}$ & $-ie(p'+p)_{\alpha}$\tabularnewline
				\hline 
			\end{tabular}
			\par\end{centering}
		\caption{Type I, II, Lepton-specific, Flipped 2HDM couplings normalized to those of the SM. Here $V=W,\,Z$.}\label{Couplings}
	\end{table}


\section{The decay $H\to h\gamma\gamma$}\label{Sec:WidthDecay}

\begin{figure}[!t]
	\centering
\includegraphics[width=0.7\textwidth]{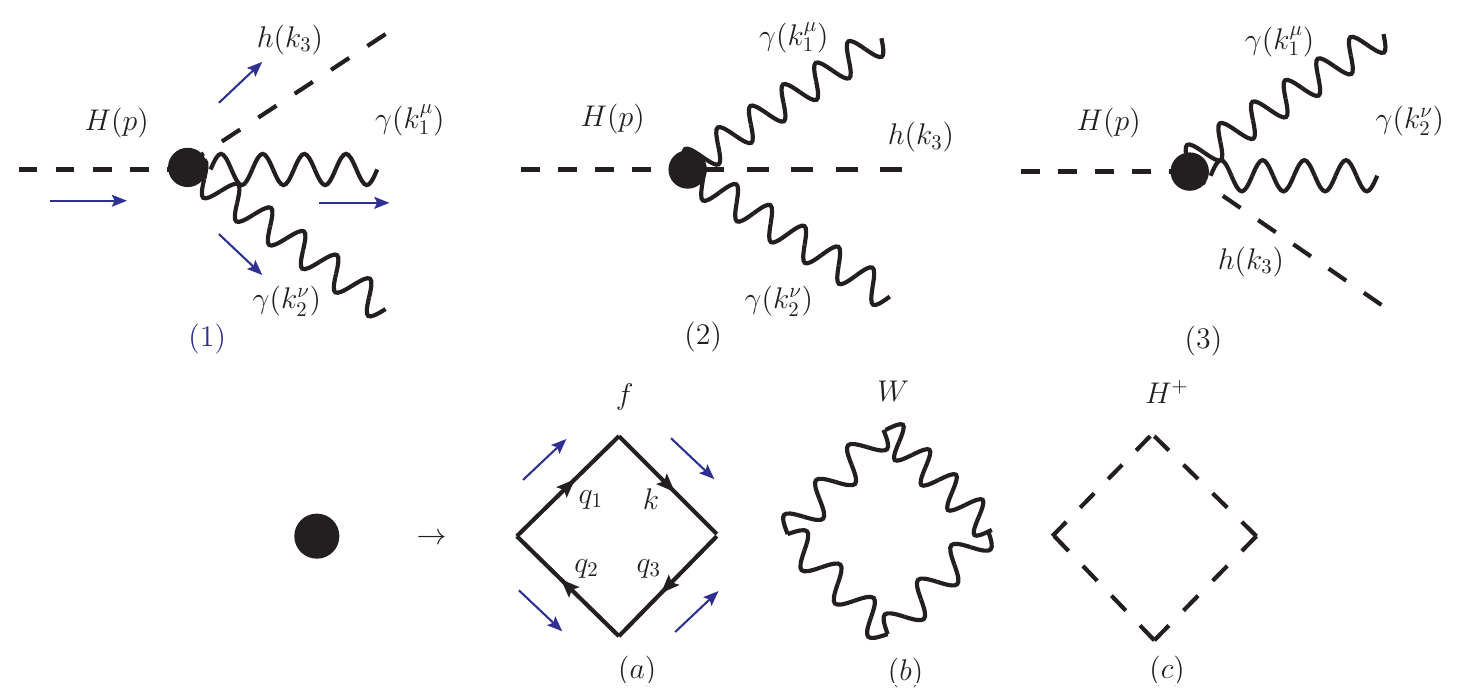}\vspace{2cm}
\includegraphics[width=0.6\textwidth]{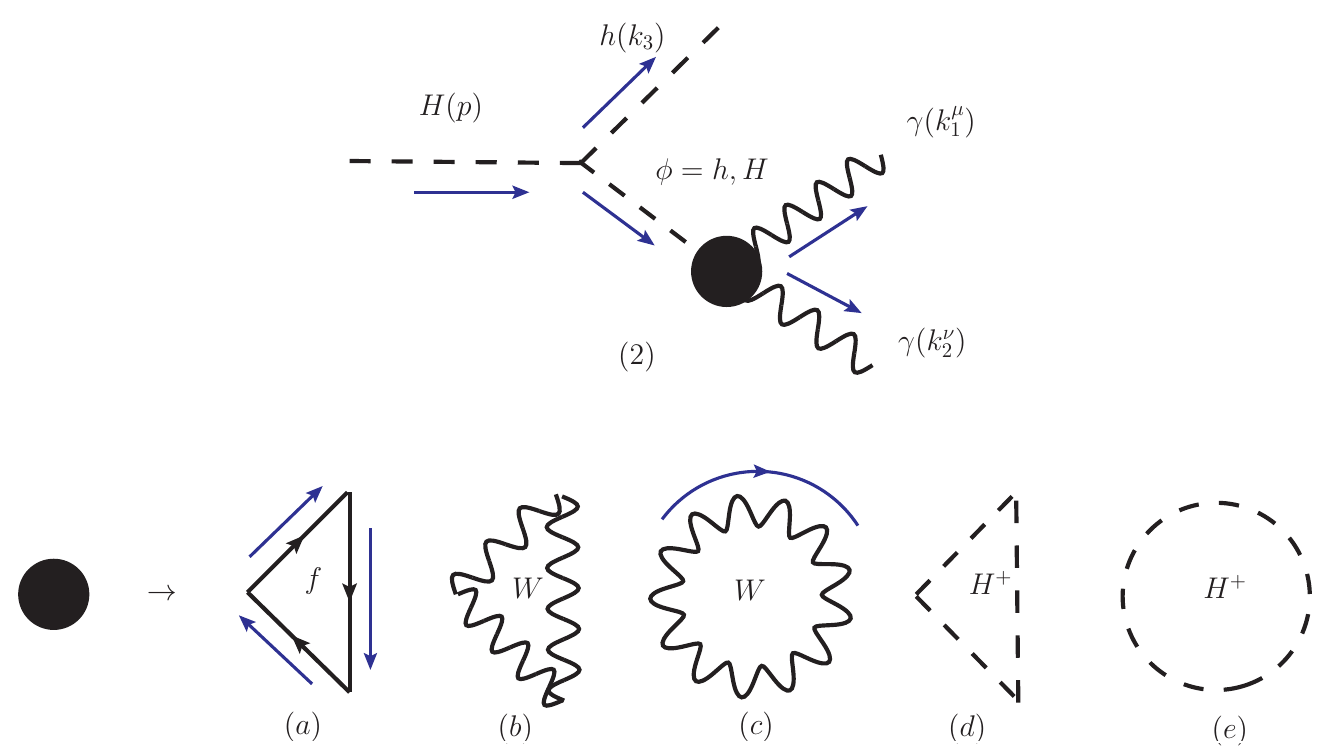}
	\caption{Box diagrams (top panel) and reducible diagrams (bottom panel) that contribute to the rare decay $H\to h\gamma\gamma$ in the 2HDM (the blue arrows indicate the momentum flow).
 Diagrams exchanging the external photon legs are not shown in the Figure. }
 \label{FeynDiagrams}
\end{figure}

In this section we calculate the rate for the rare decay $H(p)\to \gamma(k_1) \gamma(k_2) h(k_3)$, where $p$ is the momentum of the heavy Higgs, $k_1$ and $k_2$ the momenta of the photons, and $k_3$ the momentum of the SM-like Higgs.
This process is generated in the 2HDM by two kinds of Feynman diagrams: box contributions and  so-called reducible diagrams, shown in Fig.~\ref{FeynDiagrams}.

The decay rate is given by:
\begin{equation}\label{WidthDecay}
\Gamma(H\to  h\gamma\gamma)=\frac{M_H}{128\pi^3}\int_{l_{1i}}^{l_{1f}}\int_{l_{2i}}^{l_{2f}}\overline{|\mathcal{M}(H\to h\gamma\gamma)|}^2 dl_2 dl_1,	
\end{equation}
where the kinematical limits in Eq. \eqref{WidthDecay} are
\begin{align}
	l_{1i}&=2\sqrt{\mu_h},\nonumber \\
	l_{1f}&=1+2\mu_h,\nonumber\\
	l_{2i}&=\frac{1}{2}\Bigg(2-l_1-\sqrt{l_1^2-4\mu_h}\Bigg),\nonumber\\
    l_{2f}&=\frac{1}{2}\Bigg(2-l_1+\sqrt{l_1^2-4\mu_h}\Bigg),
\end{align}  
with $\mu_h=M_h^2/M_H^2$. Finally, we have calculated the invariant amplitude for the three-body process in the unitary gauge and expressed the results in terms of Passarino-Veltman functions \cite{Passarino:1978jh}.~\footnote{The algebra was carried out with the aid of \texttt{FeynCalc}\cite{Mertig:1990an}, and the result was cross-checked using
\texttt{PackageX} \cite{Patel:2015tea}.} The final result is fairly complicated, and is included in  App. \ref{amplitudes}. 

On the other hand, when $M_H>2M_h$, the rate for $H\to h\gamma\gamma$ can be approximated from the rate of the two-body decay $H\to hh$ followed by the decay of one of the Higgs bosons into two-photons, $h\to\gamma\gamma$. Under the narrow width approximation, the decay width reads
\begin{equation}\label{NarrowApprox}
	\Gamma(H\to h\gamma\gamma)=\Gamma(H\to hh)\mathcal{BR}(h\to\gamma\gamma),
	\end{equation}
where
\begin{equation}
	\Gamma(H\to hh)=\frac{G_{F}^{2}}{\sqrt{2}16\pi}\frac{M_Z^4}{M_H}\sqrt{\Bigg(1-4\frac{M_h^2}{M_H^2}\Bigg)}\Bigg(g_H^{hh}\Bigg)^2,
\end{equation}
with the (dimensionful) coupling $g_H^{hh}$ given in Appendix \ref{Acoplos}, and
\begin{equation}
\Gamma(h\to\gamma\gamma)=\frac{\alpha^2 M_h^3}{1024\pi^3 M_W^2}\Bigg|\sum_{s}A_s^{h\gamma\gamma}(\tau_s)\Bigg|^2,
\end{equation}
where the subscript $s=0,1/2,1$ refers to the spin of the charged particle circulating in the loop, and 
\begin{align}
	A_{s}^{h\gamma\gamma}=
\begin{cases}
\displaystyle{\sum_{f}\frac{2m_{W}g_{h}^{f}N_{c}Q_{f}^{2}}{m_{f}}\Big[-2\tau_{s}\big(1+(1-\tau_{s})f(\tau_{s})\big)\Big]}&{\rm for~ } s=\frac{1}{2},\\
\displaystyle{g_{hWW}\Big[2+3\tau_{W}+3\tau_{W}(2-\tau_{W})f(\tau_{W})\Big]}&{\rm for~ } s=1,\\
\displaystyle{\frac{m_{W}g_{hH^{-}H^{+}}}{M_{H^{\pm}}^{2}}\Big[\tau_{H^{\pm}}(1-\tau_{H^{\pm}}f(\tau_{H^{\pm}}))\Big]},&{\rm for~ } s=0,
 \end{cases}
\end{align}
where $N_{c}=1,3$ for leptons and quarks, respectively,
$\tau_a=4M_a^2/M_H^2$, and 
\begin{align}
	f(x)=
 \begin{cases}
        \displaystyle{\arcsin^2\Bigg(\frac{1}{\sqrt{x}}\Bigg),}& x\geq1,\\
		\displaystyle{-\frac{1}{4}\Bigg[\log\Bigg(\frac{1+\sqrt{1-x}}{1-\sqrt{1-x}}\Bigg)-i\pi\Bigg]^{2}},&x<1.
  \end{cases}
  \end{align}

We will show later that this approximation is only accurate at the ${\cal O}(20\%)$ level and that the contribution from the box diagrams in Fig.~\ref{FeynDiagrams} to the total number of events at a collider can be non-negligible. Further,
it is important to note that the contributions to the amplitude from top quarks, $W$ bosons and charged Higgses are identical for the four 2HDMs under consideration. On the other hand, the contributions from bottom quarks and tau leptons are not, as apparent from Table \ref{Couplings2}, which shows the combination of couplings that appear in the amplitude. Therefore, the measurement of the rate of $H\rightarrow h\gamma\gamma$ may be used to discriminate among the various 2HDMs. More concretely, in the alignment limit, and for the region where $\cos(\beta-\alpha)\sim 0$, $\sin\alpha\cos\alpha ~\sim -\sin\beta\cos\beta$. Therefore, one finds that for the Type I 2HDM, all couplings in Table \ref{Couplings2} are suppressed by $\tan\beta$, while for the other 2HDMs there are couplings enhanced by $\tan\beta$: the coupling to the down-type quarks and charged leptons for Type II, the coupling to the leptons for the Lepton-specific, and the coupling to the down-type quarks for the Flipped. This observation accords with our findings in the next section.

\begin{table}
		\begin{centering}
			\begin{tabular}{ccccc}
				\hline 
				Scalar part of the product \\
				of couplings & Type I & Type II & Lepton-specific & Flipped\tabularnewline
				\hline 
				\hline 
				$g_h^Ug_H^U(g_\gamma^U/q_U)^2$ & $\frac{\sin{\alpha}\cos{\alpha}}{\sin^{2}{\beta}}e^{2}$ & $\frac{\sin{\alpha}\cos{\alpha}}{\sin^{2}{\beta}}e^{2}$ & $\frac{\sin{\alpha}\cos{\alpha}}{\sin^{2}{\beta}}e^{2}$ & $\frac{\sin{\alpha}\cos{\alpha}}{\sin^{2}{\beta}}e^{2}$\tabularnewline
				\hline 
				$g_h^Dg_H^D(g_\gamma^D/q_D)^2$ & $\frac{\sin{\alpha}\cos{\alpha}}{\sin^{2}{\beta}}e^{2}$ & $-\frac{\sin{\alpha}\cos{\alpha}}{\cos^{2}{\beta}}e^{2}$ & $\frac{\sin{\alpha}\cos{\alpha}}{\sin^{2}{\beta}}e^{2}$ & $-\frac{\sin{\alpha}\cos{\alpha}}{\cos^{2}{\beta}}e^{2}$\tabularnewline
				\hline 
				$g_h^\ell g_H^\ell (g_\gamma^\ell/q_\ell)^2$ & $\frac{\sin{\alpha}\cos{\alpha}}{\sin^{2}{\beta}}e^{2}$ & $-\frac{\sin{\alpha}\cos{\alpha}}{\cos^{2}{\beta}}e^{2}$ & $-\frac{\sin{\alpha}\cos{\alpha}}{\cos^{2}{\beta}}e^{2}$ & $\frac{\sin{\alpha}\cos{\alpha}}{\sin^{2}{\beta}}e^{2}$\tabularnewline
				\hline 
			\end{tabular}
			\par\end{centering}
	\caption{Type I, II, Lepton-specific, Flipped 2HDM relevant combination of couplings entering the fermion-loop diagrams in Fig.~\ref{
 FeynDiagrams}
, illustrating the combined effect of the $D$- and $\ell$-type loops discriminating the model type.} \label{Couplings2}
	\end{table}

Let us briefly comment on the rare decay of the pseudoscalar $A\to h\gamma\gamma$. The amplitude receives contributions from the box and reducible Feynman diagrams shown in Figs. \ref{
FeynDiagrams}
($H\leftrightarrow A$) and \ref{
FeynDiagrams}(a) ($H\leftrightarrow A$ and $\phi=A$), respectively. On the other hand, the CP invariance forbids the coupling to $W^{\pm}$ or $H^{\pm}$. Thus, last row Feynman diagrams \ref{
FeynDiagrams}(b)-(e) do not contribute. Furthermore, the tree-level decay $A\to h h $ with on-shell Higgs bosons are also forbidden. As a result, the rate of  $A\to h\gamma\gamma$ is suppressed compared with $H\to h\gamma\gamma$ by up to five orders of magnitude, thus making the detection of this decay very challenging experimentally.


\section{Collider analysis}\label{Sec:Collider}
 In this section, we discuss the collider signatures of a heavy Higgs scalar $H$ decaying into SM-like Higgs, $h$, and two photons at Run 3 of the LHC, as well as at the HL-LHC. Our study considers a center-of-mass energy of $\sqrt{s}=14 $ TeV for both colliders, and luminosities in the range 300-3000 fb$^{-1}$.
  
\begin{figure}[!t]
\begin{center}
\includegraphics[width=0.45\textwidth]{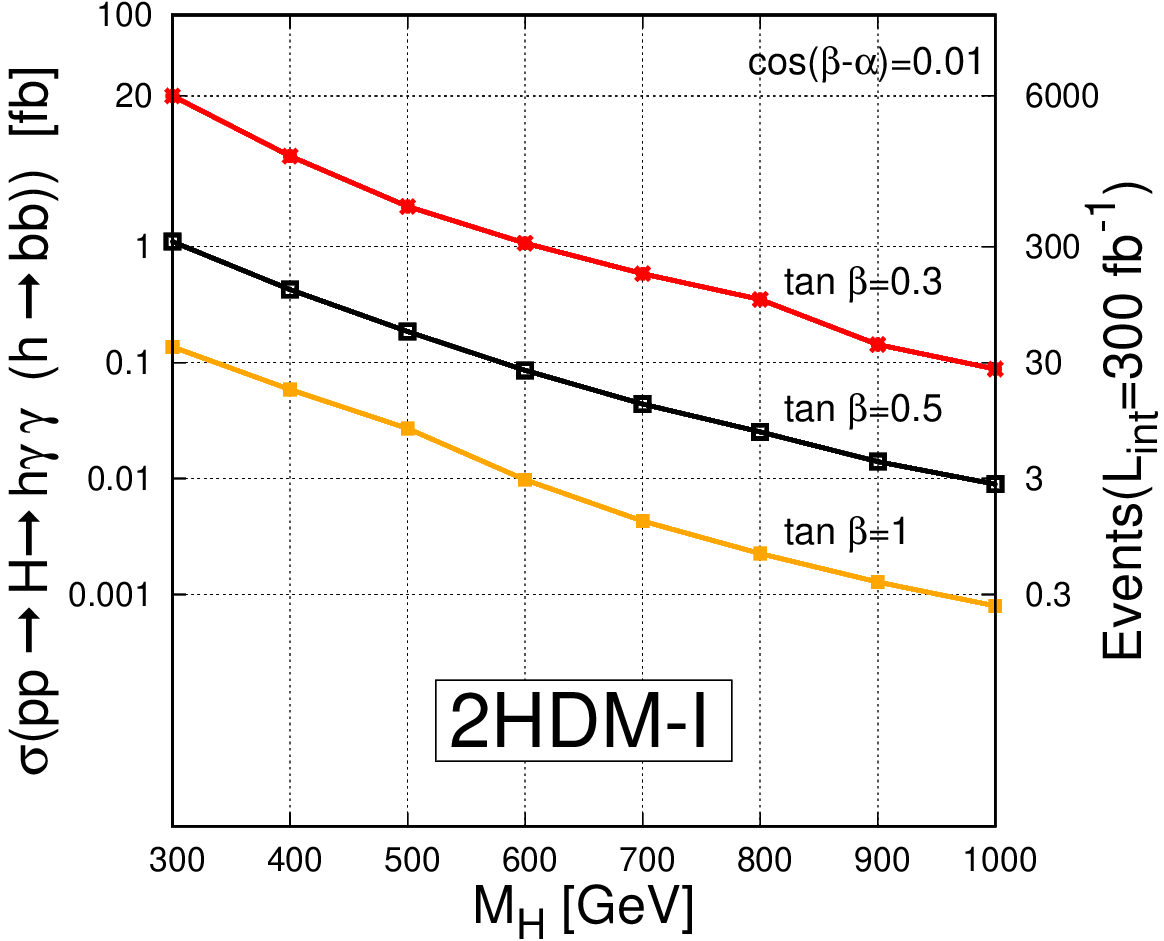}
\includegraphics[width=0.45\textwidth]{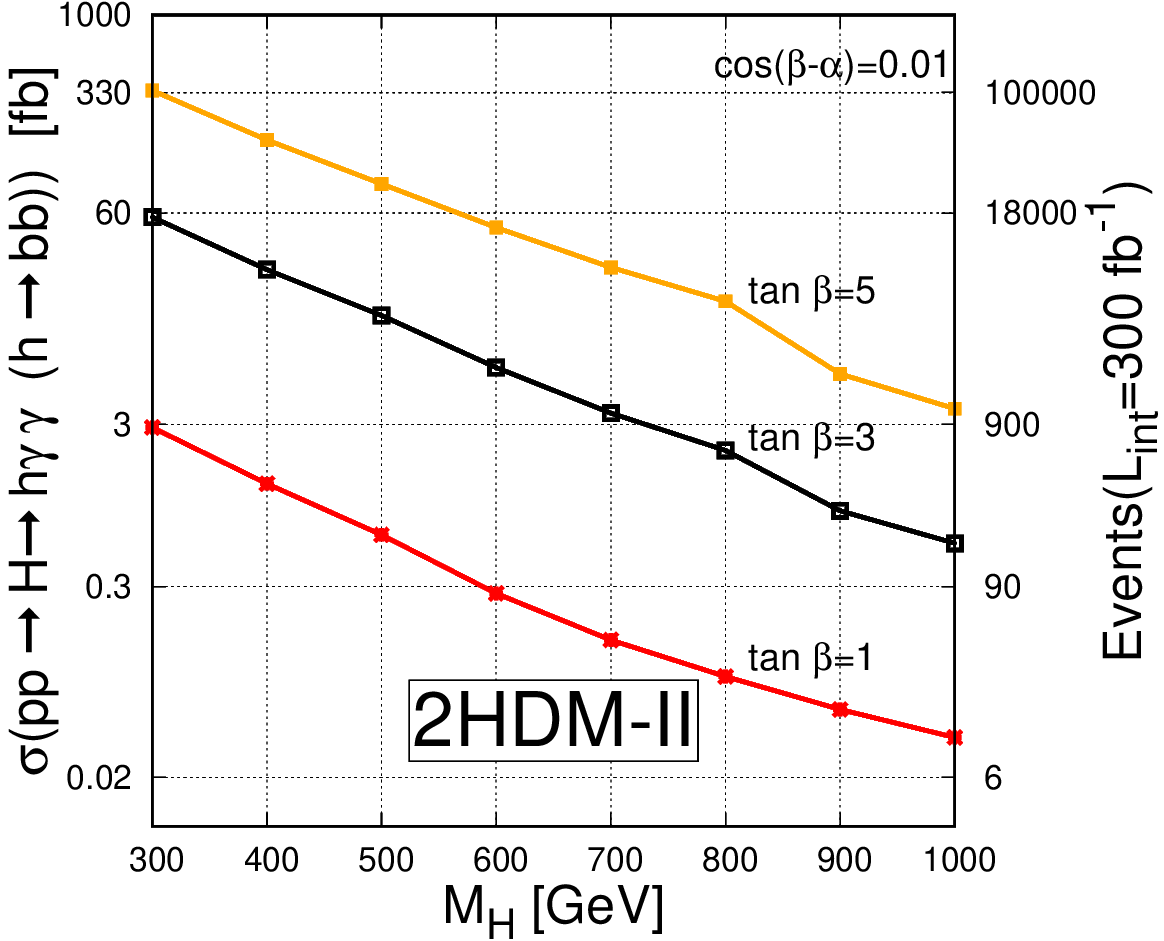}
\\\includegraphics[width=0.45\textwidth]{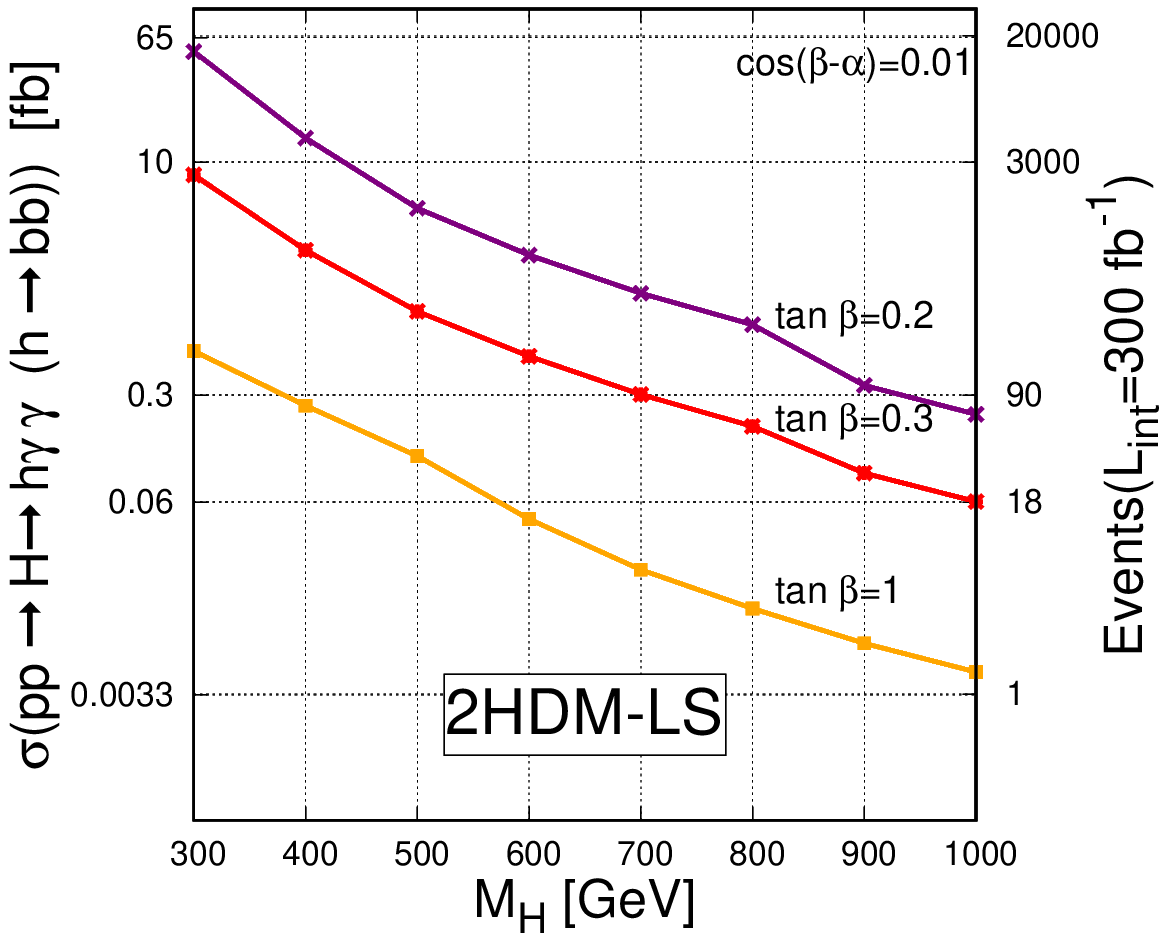}
\includegraphics[width=0.45\textwidth]{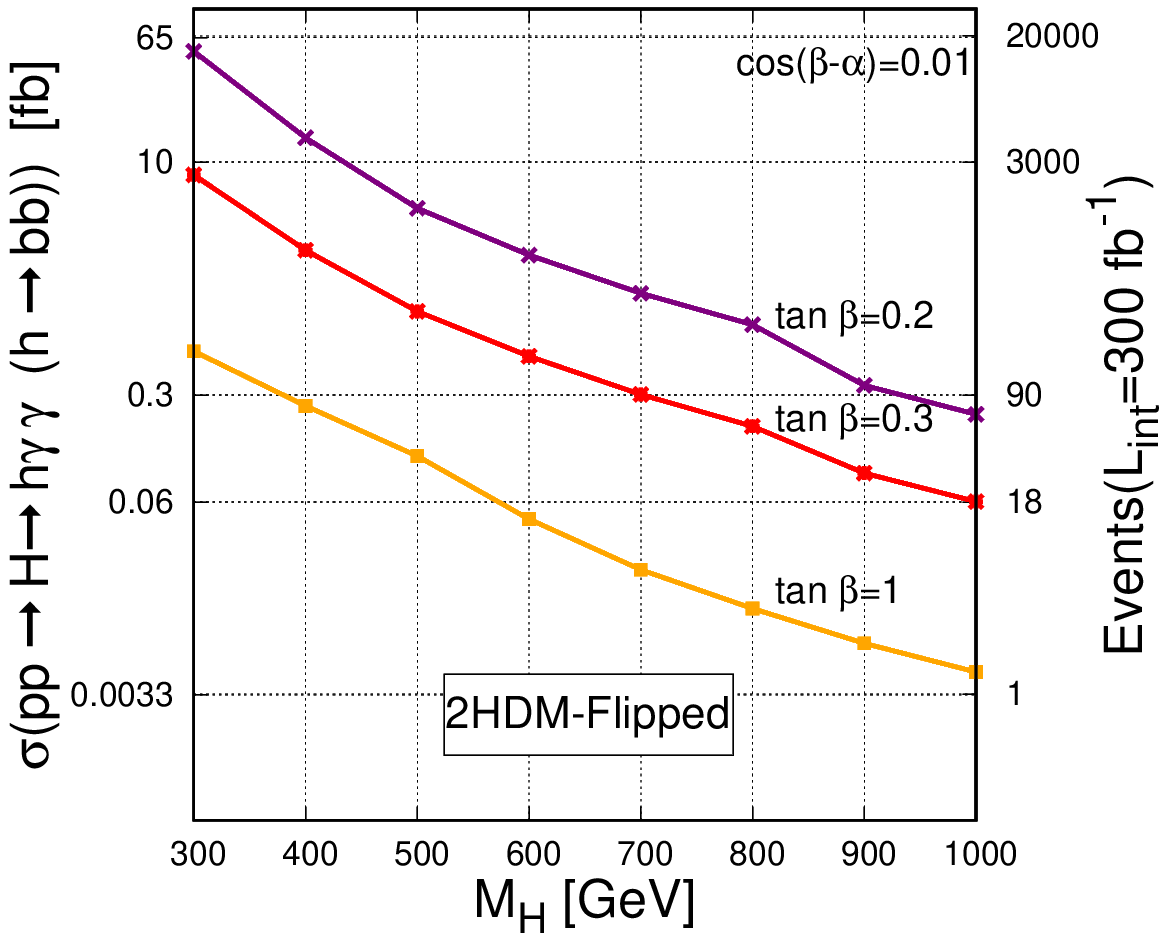}
\end{center}
\caption{Production cross-section and number of events for the process $H\to h \gamma\gamma$ followed by $h\to b\bar{b}$.}
\label{XSallModels}
\end{figure}

 We focus on the search for the final state $\gamma\gamma b\bar{b}$,  emerging from the resonant production of a heavy Higgs boson in a proton-proton collision. We show in Fig. \ref{XSallModels} the cross-section -including the contribution of the pure two-body decay- for the process $pp \to H\to h \gamma\gamma ~(h \to b \bar{b})$ as a function of $M_H$, for different values of $\tan\beta$, and assuming $\cos(\beta-\alpha)=0.1$ for the 2HDM-I and $\cos(\beta-\alpha)=0.01$ for the rest. We also show on the right axis the number of events produced for an integrated luminosity of 300 fb$^{-1}$, \textit{i.e.}, the projected goal of the LHC Run 3. 

\begin{figure}[!t]
\begin{center}
	\subfigure[]{\includegraphics[scale=0.45]{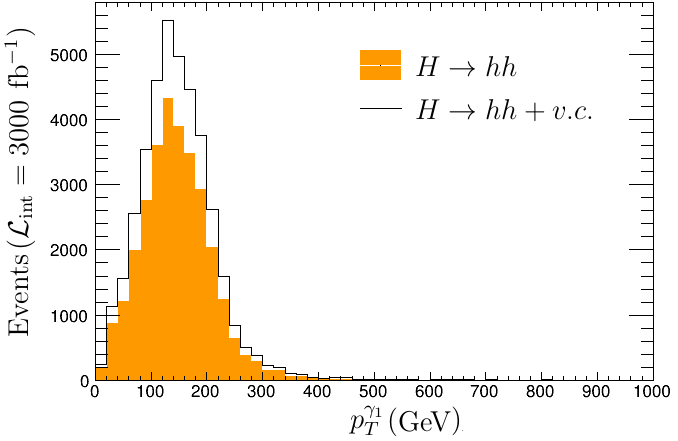}}
 \subfigure[]{\includegraphics[scale=0.45]{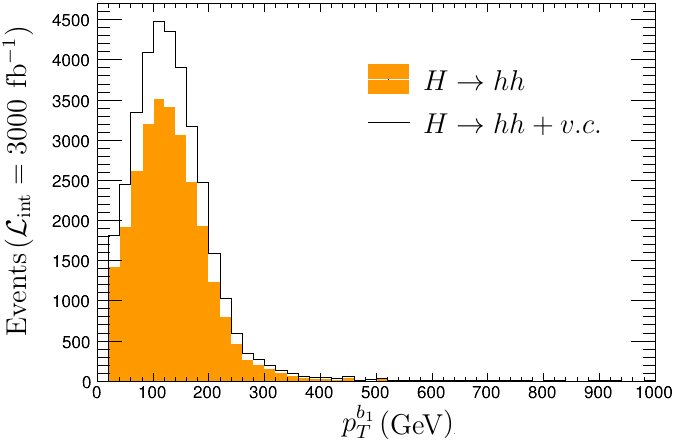}}
 \subfigure[]{\includegraphics[scale=0.45]{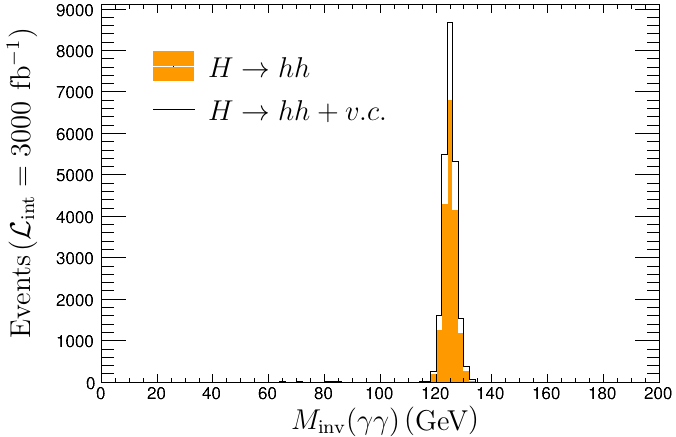}}
 \subfigure[]{\includegraphics[scale=0.45]{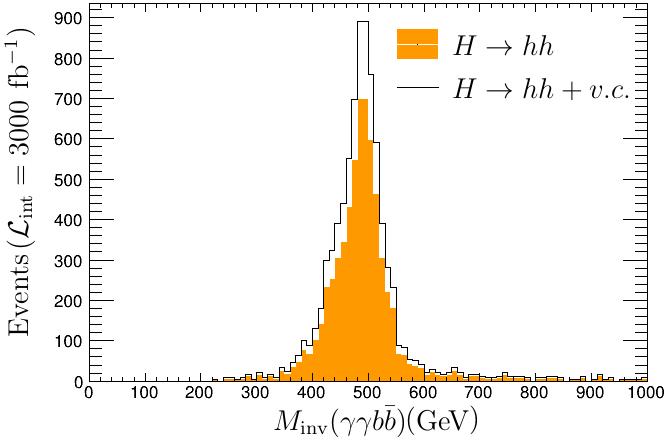}}
  \end{center}
	\caption{Number of events distributions (for $\mathcal{L}_{int}=3000$ fb$^{-1}$) as functions of: (a) Transverse momentum of the photon, (b) Transverse momentum of the $b$-jet, (c) Invariant mass of the two photons, and (d) Invariant mass of the $\gamma\gamma b\bar{b}$ system for $M_H=500$ GeV. Index $1$ represents the leading photon and leading $b-jet$. The orange distribution corresponds to the events produced by the process $H\to hh$, while the solid black distribution stands for the contribution of $H\to hh$ plus virtual contributions (v.c.). We note that the ratio between the black and orange histograms is not constant along the $x$ variable in these plots.}  
 \label{distributions}
\end{figure}

  It should be noted that the dominant contribution has its source in the cascade decay $H\to hh\to \gamma\gamma b\bar{b}$. However, we find that virtual contributions are necessary, as they would show as an excess of events of $\mathcal{O}(20\%)$ relative to the two-body process ($hh$). Thus, neglecting the $h\gamma\gamma$ subprocess would amount to a sizable unitarity violation. Different event distributions abound on this point, as displayed in Fig.~\ref{distributions}.

Potential SM backgrounds come mainly from $h+X$, where
$X=Z,\,b\bar{b},\,t\bar{t}$, as well as non-Higgs processes including $t\bar{t}$ and $t\bar{t}\gamma$ (leptons arise in these processes that may fake photons) and $b\bar{b}\gamma\gamma$, $c\bar{c}\gamma\gamma$, $jj\gamma\gamma$ (here $c-$jets and light-jets may mimic $b$-jets). In addition, we also include relevant reducible backgrounds such as $b\bar{b}j\gamma$, $c\bar{c}j\gamma$ and $b\bar{b}jj$, where $c$-jets may appear as $b$-jets and a light-jet may fake a photon. The fake rate of a light-jet $j$ into a photon depends on the transverse momentum of the jet $p^j_T$ \cite{ATLAS:2013kpx} as $9.3\times 10^{-3}\exp (-p^i_T/27.5\, \rm GeV)$. As far as the $c$-jet is concerned, it can be misidentified as a $b$-jet with a rate of 3.5\% while a light-jet mimics a $b$-jet with a probability of 0.135\% \cite{CMS:2017wtu}. Table \ref{tab:csBG} summarizes the SM background processes, along with the corresponding production cross sections.
\begin{table}[t!]
	\begin{center}\scalebox{1.0}{
			\begin{tabular}{|cc|}
				\hline
				SM backgrounds & Cross section [pb]  \\
				\hline
				\rule{0pt}{1ex}
				$ pp \to b \bar{b} \gamma \gamma  $&    $5.14$    \\
				$ pp \to Zh \, ( Z\to b \bar{b}, h\to \gamma \gamma)  $&       $1.41 \times 10^{-4}$ \\
				\hline
				$ pp \rightarrow b \bar{b} j \gamma  $ &    $9750$    \\
				$ pp \rightarrow b \bar{b} j j  $ &    $7.74\times 10^6$    \\
				($j$ mimics a photon)&\\
				\hline
				$ pp \rightarrow c \bar{c} \gamma \gamma  $ &    $6.66 $   \\
				$ pp \rightarrow c \bar{c} j \gamma  $ &    $ 2636 $  \\
				($c$ appears as $b$-tagged jet,&              \\
				$j$ mimics a photon)&        \\
				\hline
				$ pp \rightarrow jj \gamma \gamma  $ &    $ 112.40 $  \\
				($j$ appears as $b$-tagged jet)&              \\
				\hline
				$ pp \rightarrow t \bar{t}\, (t \to \bar{\ell} {\nu}_\ell b,  \bar{t}\to \ell \bar{\nu}_\ell \bar{b} )$ &   $27.35$     \\
				$ pp \rightarrow t \bar{t}\, (t \to jj b,  \bar{t}\to jj \bar{b} )$ &   $ 245.50   $ \\
				($\ell,j$ mimic a photon)&     \\
				\hline
		\end{tabular}}
	\end{center}
	\caption{The cross sections for the most relevant SM background processes.}
	\label{tab:csBG}
\end{table}

 Insofar as our computation scheme is concerned, we used $\texttt{LanHEP}$~\cite{Semenov:2014rea} to build the 2HDM and produce the \texttt{UFO} files~\cite{Degrande:2011ua}~\footnote{We modelled the loop-induced 
contributions using an effective vertex and checked the validity of our approach.}, subsequently we generated parton-level events for both the signal and the SM background processes using $\texttt{MadGraph5}$ \cite{Alwall:2014hca} and we performed shower and hadronization with $\texttt{Pythia8}$ \cite{Sjostrand:2019zhc}. The detector response has been emulated using \texttt{MadAnalysis}~\cite{Conte:2012fm} in its reconstruction level\footnote{We use the \texttt{.hepmc} folder generated by \texttt{Pythia8} within \texttt{MadAnalysis}, then we invoke the \texttt{HL-LHC}~\cite{HL-LHC_card} configuration card to emulate the detector response.}. In the calculation of the cross section we employ the $\texttt{NN23LO1}$ PDF set \cite{NNPDF:2017mvq}.  As far as the jet reconstruction is concerned, the jet finding package $\texttt{FastJet}$ \cite{Cacciari:2011ma} and the anti-$k_T$ algorithm \cite{Cacciari:2008gp} were used\footnote{This was done through the interface of \texttt{MadAnalysis.}}.

\subsection*{b-jet identification}
\begin{figure}[!t]
\begin{center}
	\subfigure[]{\includegraphics[scale=0.37]{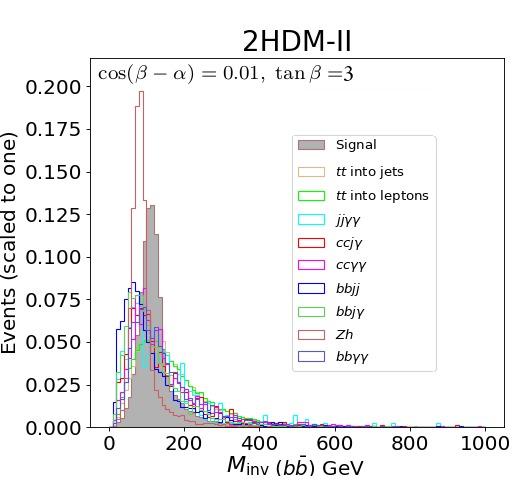}}
 \subfigure[]{\includegraphics[scale=0.3]{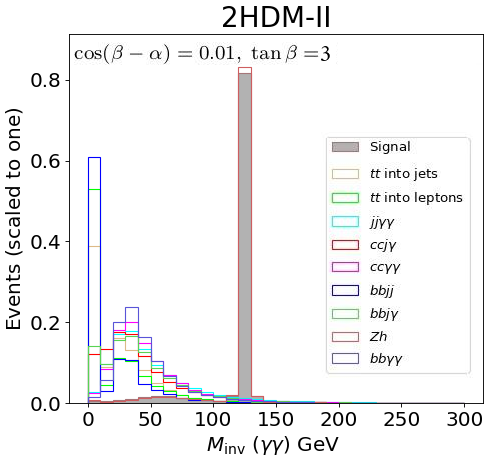}}
 \subfigure[]{\includegraphics[scale=0.37]{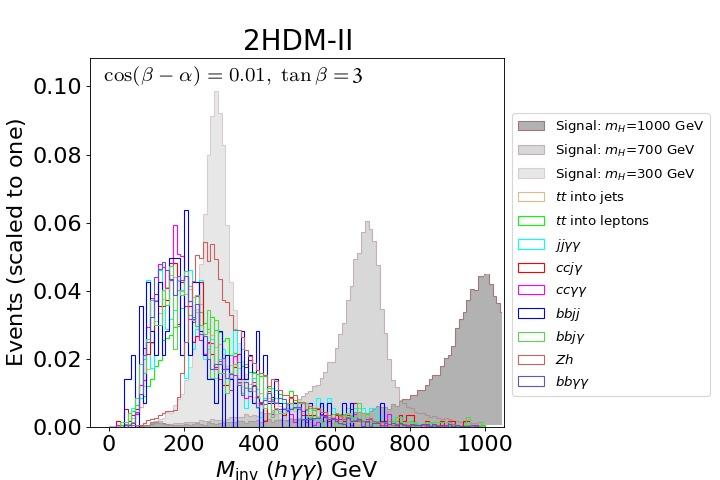}}
 \end{center}
	\caption{Invariant mass distributions for (a) $M_{\rm inv}(b\bar{b})$, (b) $M_{\rm inv}(\gamma\gamma)$, and (c) $M_{\rm inv} (h\gamma\gamma)$ in the 2HDM-II scenario with $\cos(\beta-\alpha)=0.01$, $\tan\beta=3$ and $M_A=M_{H^\pm}=M_H$.}
 \label{InvMasses}
\end{figure}

	The $b$-tagging of jets produced from the fragmentation and hadronization of bottom quarks plays a fundamental role in separating the signal from the background processes, which involve gluons, light-flavor jets ($u,\,d,\,s$) and c-quark fragmentation. To overcome this problem, we use the \texttt{FastJet} package~\cite{Cacciari:2011ma} (via \texttt{MadAnalysis}~\cite{Conte:2012fm}) and invoke the anti-$k_T$  algorithm~\cite{Cacciari:2008gp}. We also include the $b$-tagging efficiency, $\epsilon_b=90\%$. The probability that a $c-$jet or any other light-jet $j$ is mistagged as a $b-$jet are $\epsilon_c=5\%$~\cite{ATLAS:2023gog} and $\epsilon_j=1\%$, respectively. To improve the $h\to \bar{b}b$ signal discrimination against the non-resonant background and to obtain a better separation of the Higgs boson from the $Z$ boson, the invariant mass $M_{\rm inv}(\bar{b}b)$ plays a key role. From an experimental point of view, the jet energy calibration is performed as a function of the jet-$p_T$ and $\eta$. Typical jet energy resolutions are of order of $15\%$ for $p_T\sim 35$ GeV.
Based on the above, we assume the following kinematic cuts:
\begin{enumerate}
	\item We identify $b$-jets and photons by imposing the transverse momentum cuts $p_{T}\left(\gamma\right)>35$ GeV and $p_{T}\left(b\right)>40$ GeV.
    \item We require the angular separation between the $b$-jets to be $0.4<\Delta R<2.8$ and that the angular separation between  each photon pair be $\Delta R\equiv \sqrt{\Delta\phi^{2}+\Delta\eta^{2}}<3$, 
	\item We impose a cut on the invariant masses $M_{\rm inv}(\bar{b}b)$ and $M_{\rm inv}(\gamma\gamma)$ in the ranges 105 GeV $<M_{\rm inv}(b\bar{b})<$140 GeV and 122.5 GeV $<M_{\rm inv}(\gamma\gamma)<$128.5 GeV, respectively,
    \item Finally, the cuts on $M_H-50$ GeV $<M_{\rm inv} (h\gamma\gamma)<M_H+50$ GeV are also required.

    These requirements efficiently removes background events, as apparent from Figure \ref{InvMasses}, which shows the number of events from different processes (scaled to one) versus invariant mass distributions for representative choices of parameters in the 2HDM-II; we obtain analogous results for other 2HDMs.
\end{enumerate}

\subsection*{Signal significance}
Motivated by the analysis performed in Sec.~\ref{Sec:Model}, we consider the particular case of the alignment limit, \textit{i.e.}, $\cos(\beta-\alpha)\to 0$. It should be noted that for the 2HDM-II, LS and Flipped, a tail-shaped region survives all the constraints, both experimental and theoretical, as shown in Fig.~\ref{AllConstraints}. However, we are cautious in that case because, as suggested in~\cite{Arroyo-Urena:2020fkt}, such a region could be excluded by the so-called \textit{D\'iaz-Cruz} angle. Thus, we define three realistic benchmark points (BMP) on which the predictions below are based:
\begin{itemize}
    \item BMP1: $\cos(\beta-\alpha)=0.01$, $\tan\beta=0.3$ for 2HDM-I, LS and Flipped, and $\tan\beta=3$ for 2HDM-II.
    \item BMP2: $\cos(\beta-\alpha)=0.01$, $\tan\beta=0.2$ for 2HDM-I, LS and Flipped, and $\tan\beta=5$ for 2HDM-II.
    \item BMP3: $\cos(\beta-\alpha)=0.01$, $\tan\beta=1$ for all versions.
\end{itemize}
We show in Figs. \ref{Significance_2HDM-type_I}-\ref{Significance_2HDM-type_Flipped} (2HDM-I, 2HDM-II, 2HDM-Flipped, respectively~\footnote{The results for 2HDM-LS differ from those for 2HDM-Flipped by a factor of $\sim 1.05$, so we do not present this case.}) the signal significance as a function of $M_{H}$ after applying the kinematic cuts, for different values of the integrated luminosities $\mathcal{L}_{\rm int}$, and for the benchmark points shown in the inset. We evaluate the signal significance using the relation $N_S/\sqrt{N_S+N_B+(\kappa N_B)^2}$, where $N_S$, $N_B$ and $\kappa$ are the number of signal events, number of background events and the percentage of systematic uncertainty~\cite{Cowan:2012gra}, respectively. 
These figures demonstrate the LHC discovery potential for signals of the rare process $H\rightarrow h(b\bar{b})\gamma\gamma$. The prospects for discovery are especially promising for the 2HDM-II (Fig.~\ref{Significance_2HDM-type_II}). For BMP2 (BMP1), even a luminosity of $300\, {\rm fb}^{-1}$ would lead to signals with $\gtrsim 3\sigma$ if $M_H\lesssim 600$ ($M_H\lesssim 430$) GeV. Moreover, it could be possible to reach significances $\gtrsim 5\sigma$ for $M_H\lesssim 720$ ($M_H\lesssim 560$) GeV with $1000\,\ifb$ for BMP2 (BMP1), and $M_H\lesssim 940$ ($M_H\lesssim 840$) GeV with $3000\ifb$ for BMP2 (BMP1). For the 2HDM-Flipped, the sensitivity is somewhat poorer, requiring 
3000 $\ifb$ to reach a $5\sigma$ significance if $M_H\lesssim 650$ ($M_H\lesssim 440$) GeV for BMP2 (BMP1). The worst sensitivity occurs in the 2HDM-I, where it will be challenging to reach a $5\sigma$ sensitivity, even if the integrated luminosity is $\sim 3000\,\ifb$, although a $3\sigma$ signal significance could be reached in $M_H\lesssim 560$ ($M_H\lesssim 700$) GeV for BMP2 (BMP1). BMP3 is the least favored in all cases, even for an integrated luminosity of $3000\,\ifb$. In all signal significances reported above, a systematic uncertainty $\kappa=5\%$ has been considered.

\begin{figure}[!htb]
\begin{center}
\subfigure[]{\includegraphics[width=0.325\textwidth,angle=0]{
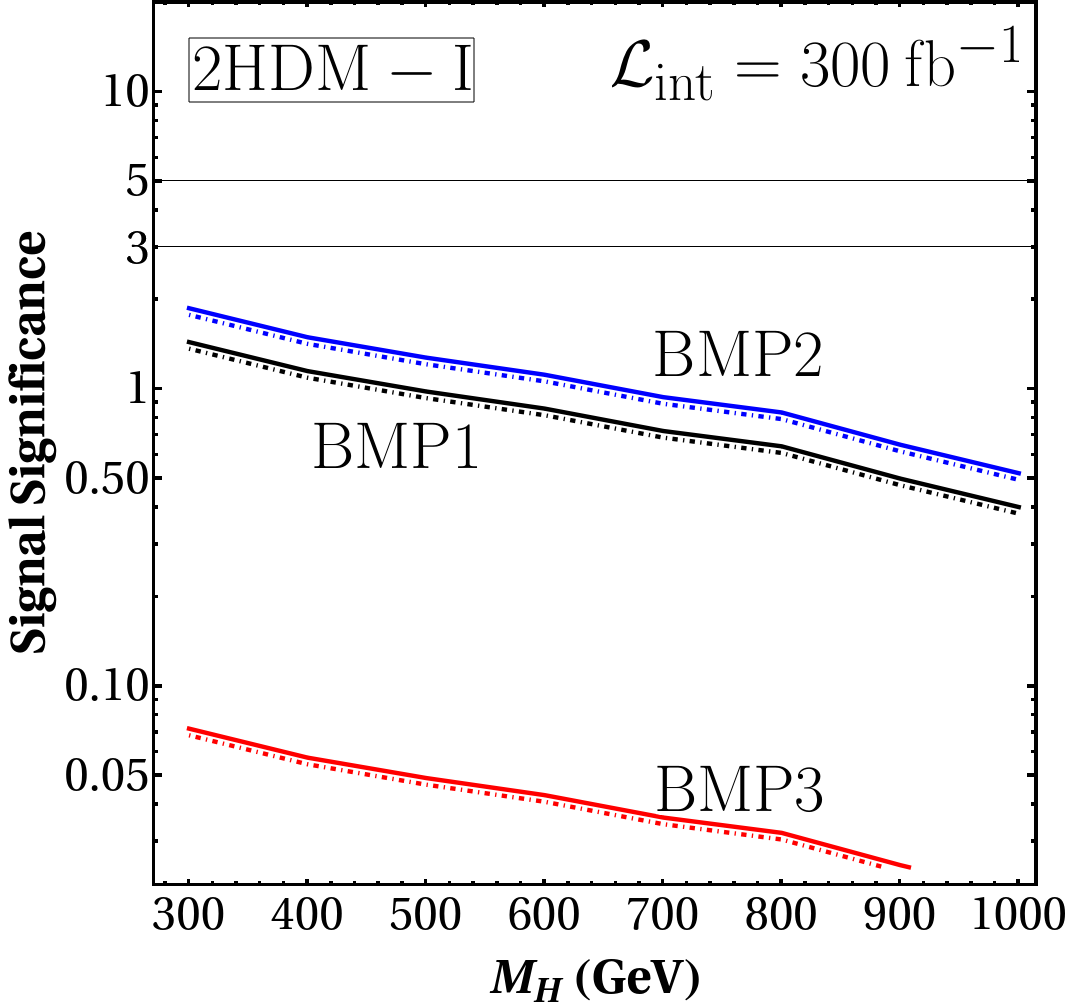}}
\subfigure[]{\includegraphics[width=0.325\textwidth,angle=0]{
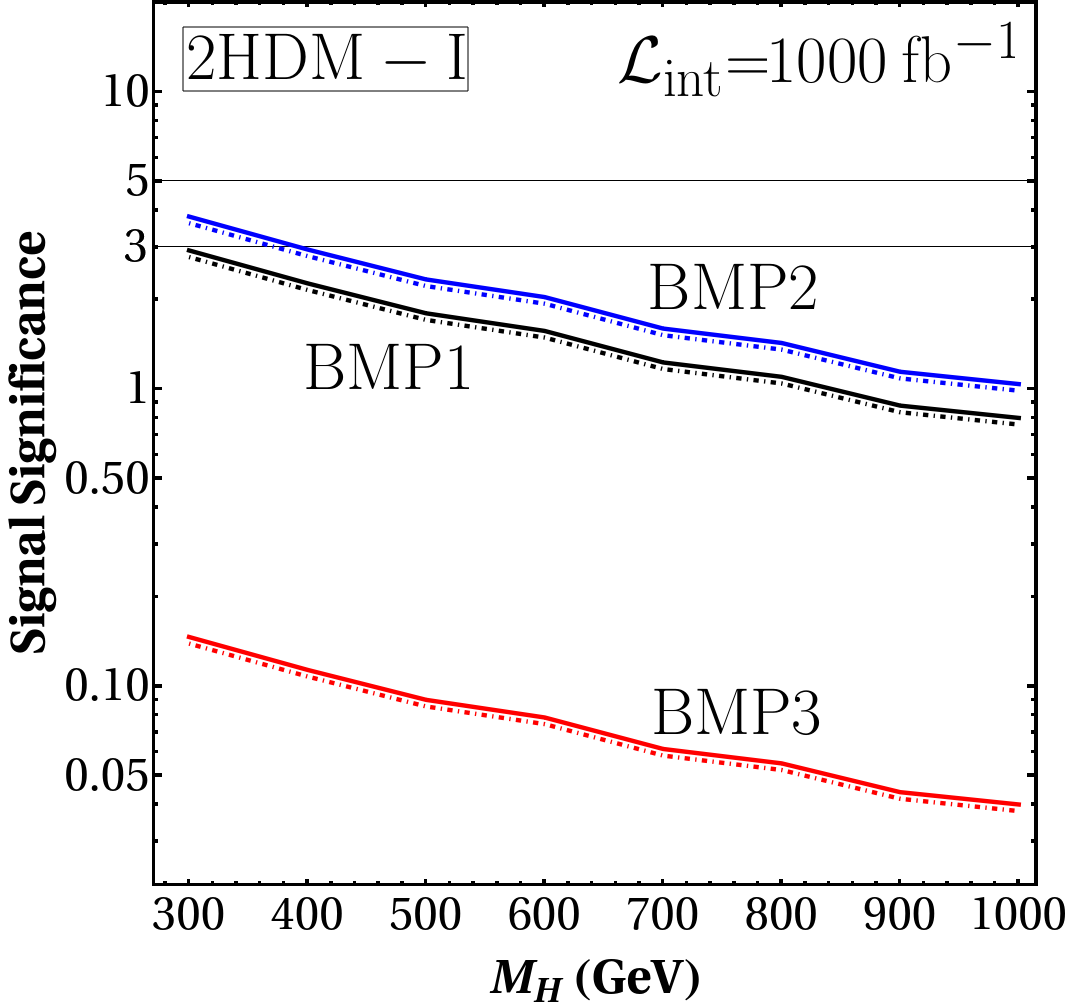}}
\subfigure[]{\includegraphics[width=0.325\textwidth,angle=0]{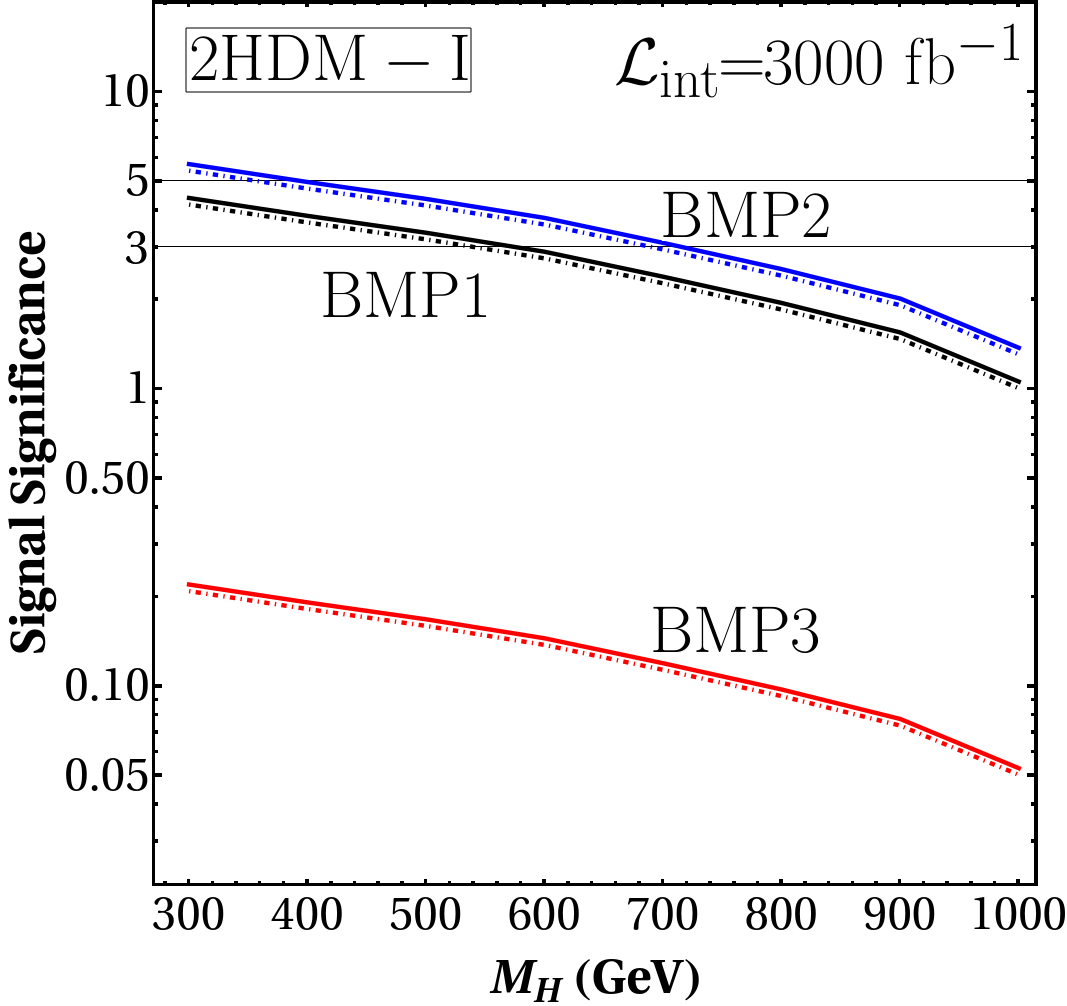}}
\end{center}
\caption{Signal significance expected at the LHC for the 2HDM-I with an integrated luminosity of (a) 300, (b) 1000 and (c) 3000 fb$^{-1}$ for the process $H\to h \gamma\gamma$ (followed by $h\to b\bar{b})$, for the three benchmark points defined in the main text. Solid lines stand for the signal significance without systematic uncertainties, while the dotdashed lines consider a systematic uncertainty of $\kappa=5\%$.}
\label{Significance_2HDM-type_I}
\end{figure}
\begin{figure}[!htb]
\begin{center}
\subfigure[]{\includegraphics[width=0.325\textwidth,angle=0]{
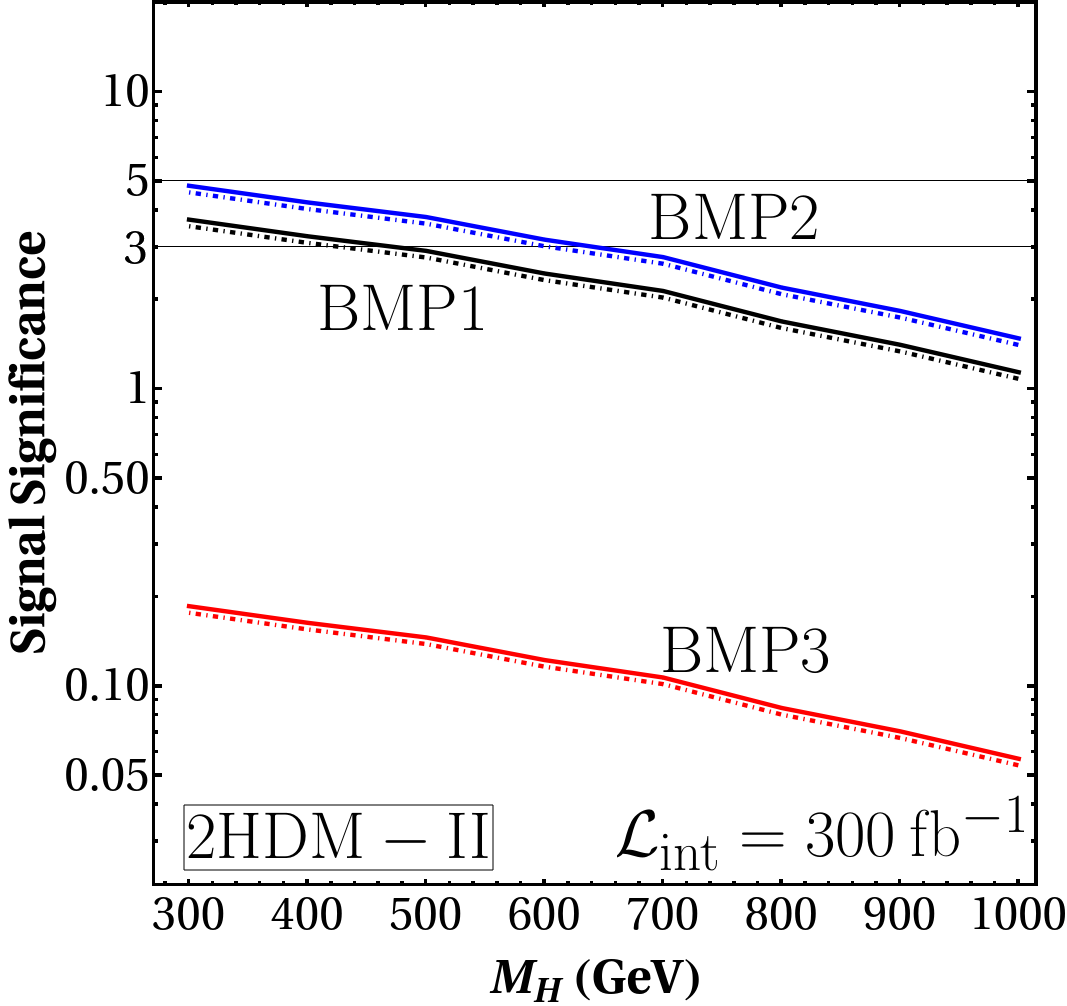}}
\subfigure[]{\includegraphics[width=0.325\textwidth,angle=0]{
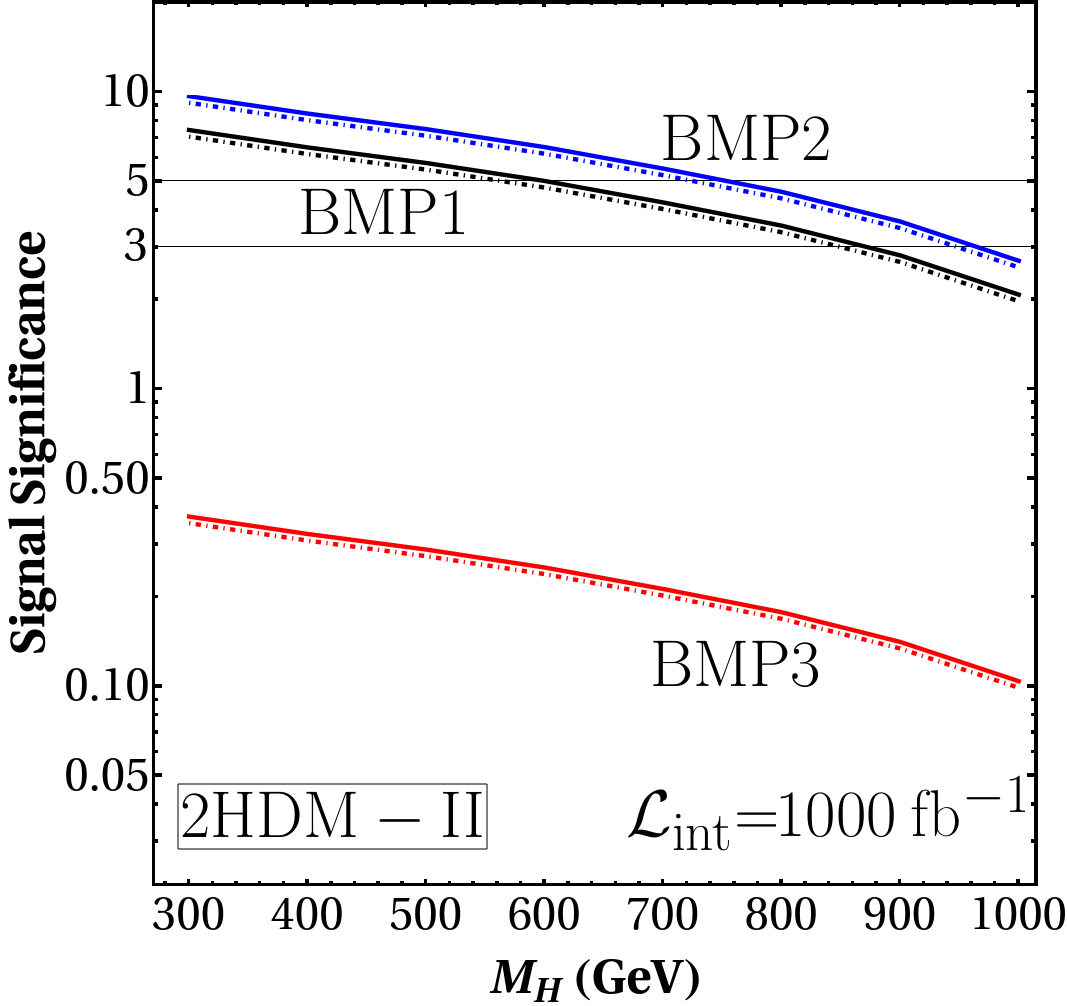}}
\subfigure[]{\includegraphics[width=0.325\textwidth,angle=0]{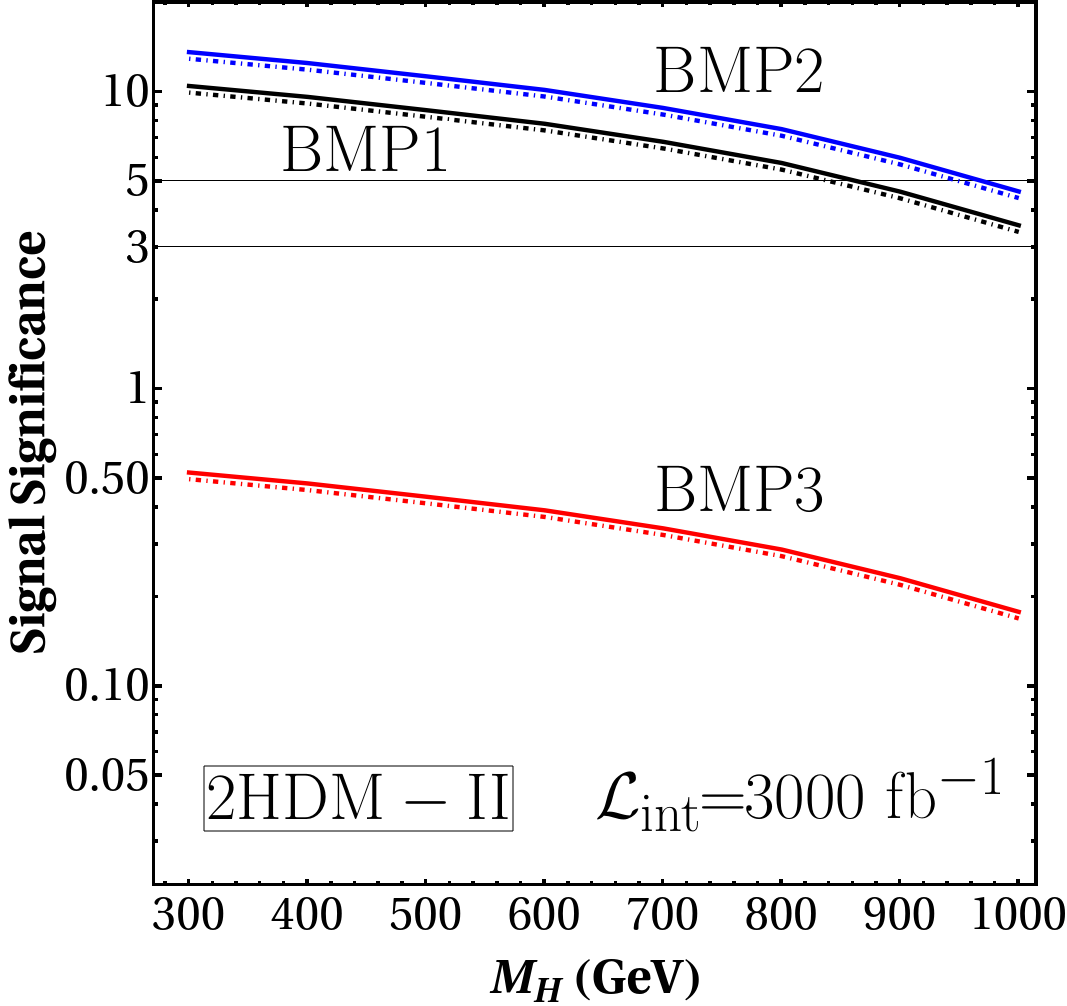}}
\end{center}
\caption{Signal significance expected at the LHC for the 2HDM-II with an integrated luminosity of (a) 300, (b) 1000 and (c) 3000 fb$^{-1}$ for the process $H\to h \gamma\gamma$ (followed by $h\to b\bar{b})$, for the three benchmark points defined in the main text. Solid lines stand for the signal significance without systematic uncertainties, while the dotdashed lines consider a systematic uncertainty of $\kappa=5\%$.}
\label{Significance_2HDM-type_II}
\end{figure}

\begin{figure}[!htb]
\begin{center}
\subfigure[]{\includegraphics[width=0.325\textwidth,angle=0]{
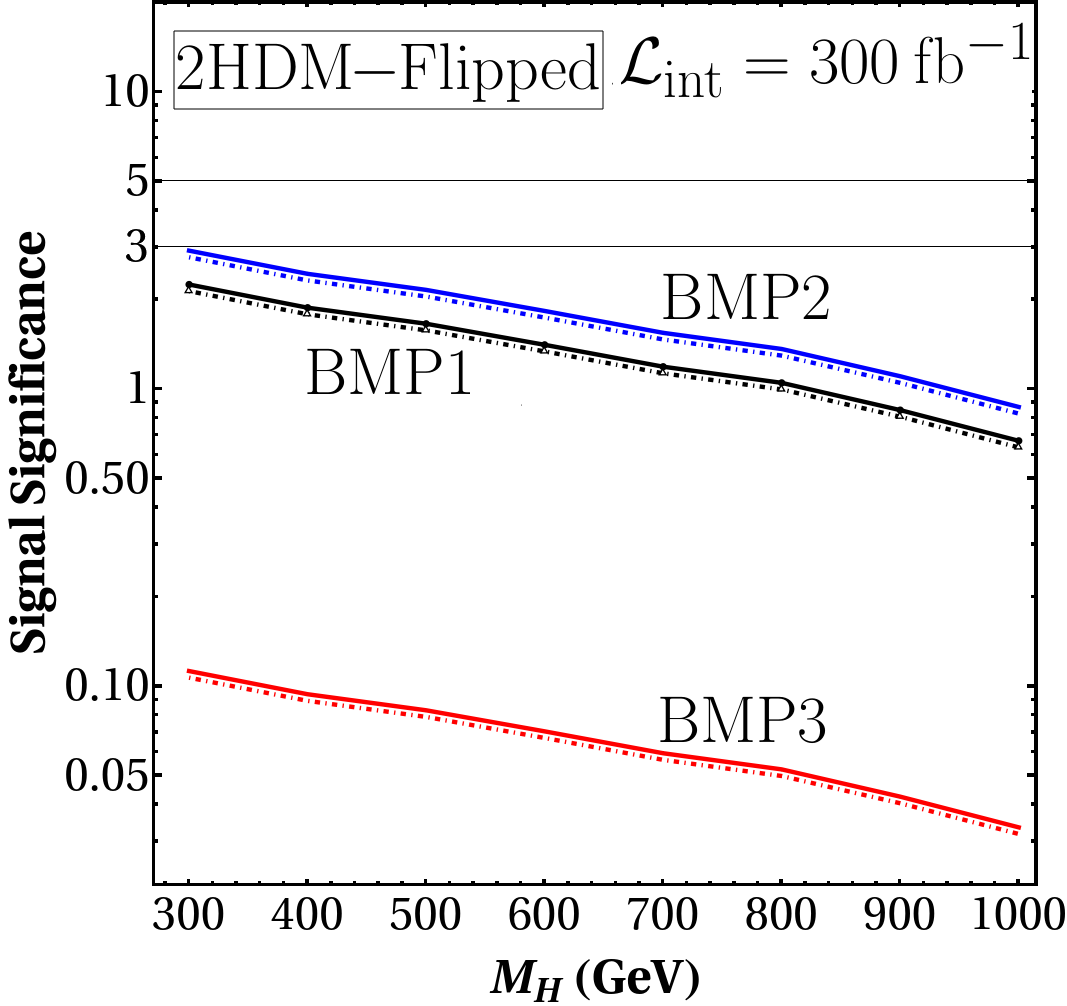}}
\subfigure[]{\includegraphics[width=0.325\textwidth,angle=0]{
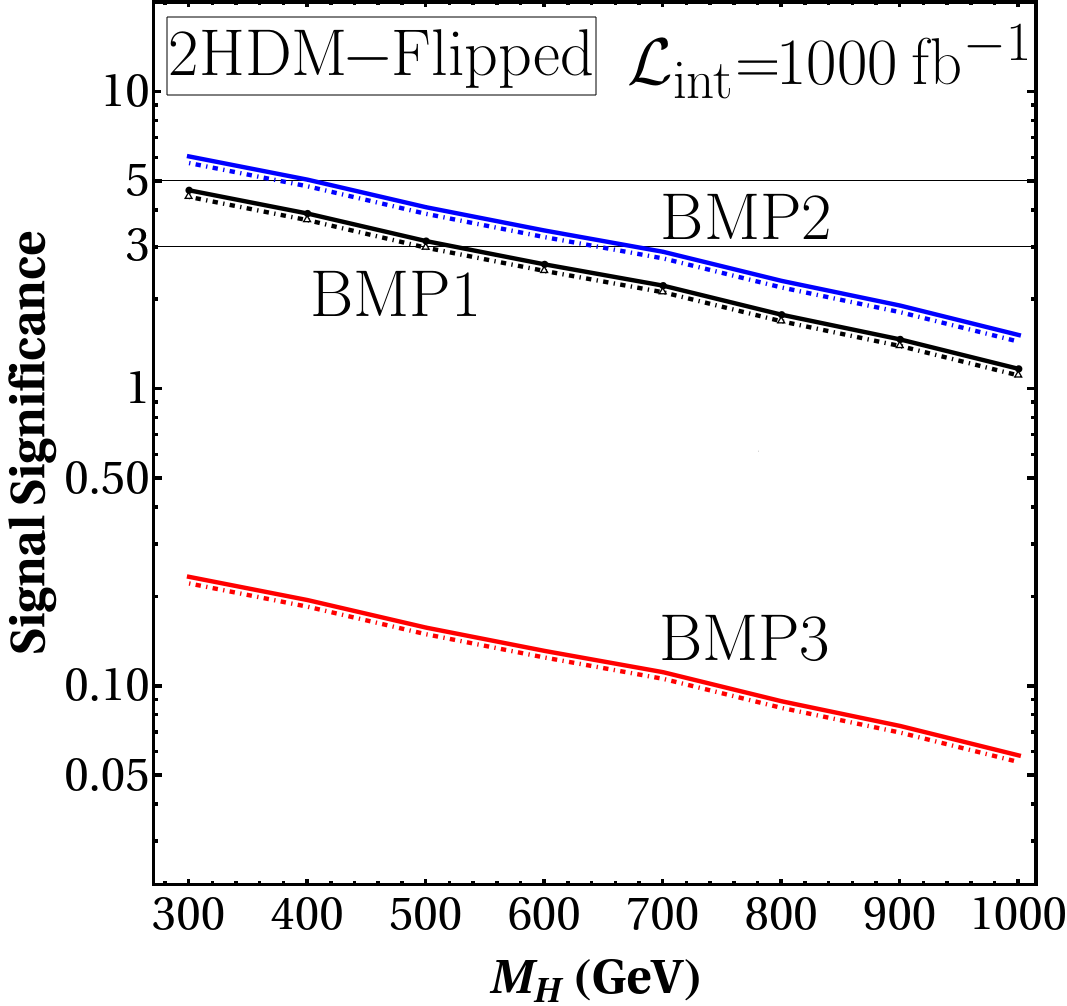}}
\subfigure[]{\includegraphics[width=0.325\textwidth,angle=0]{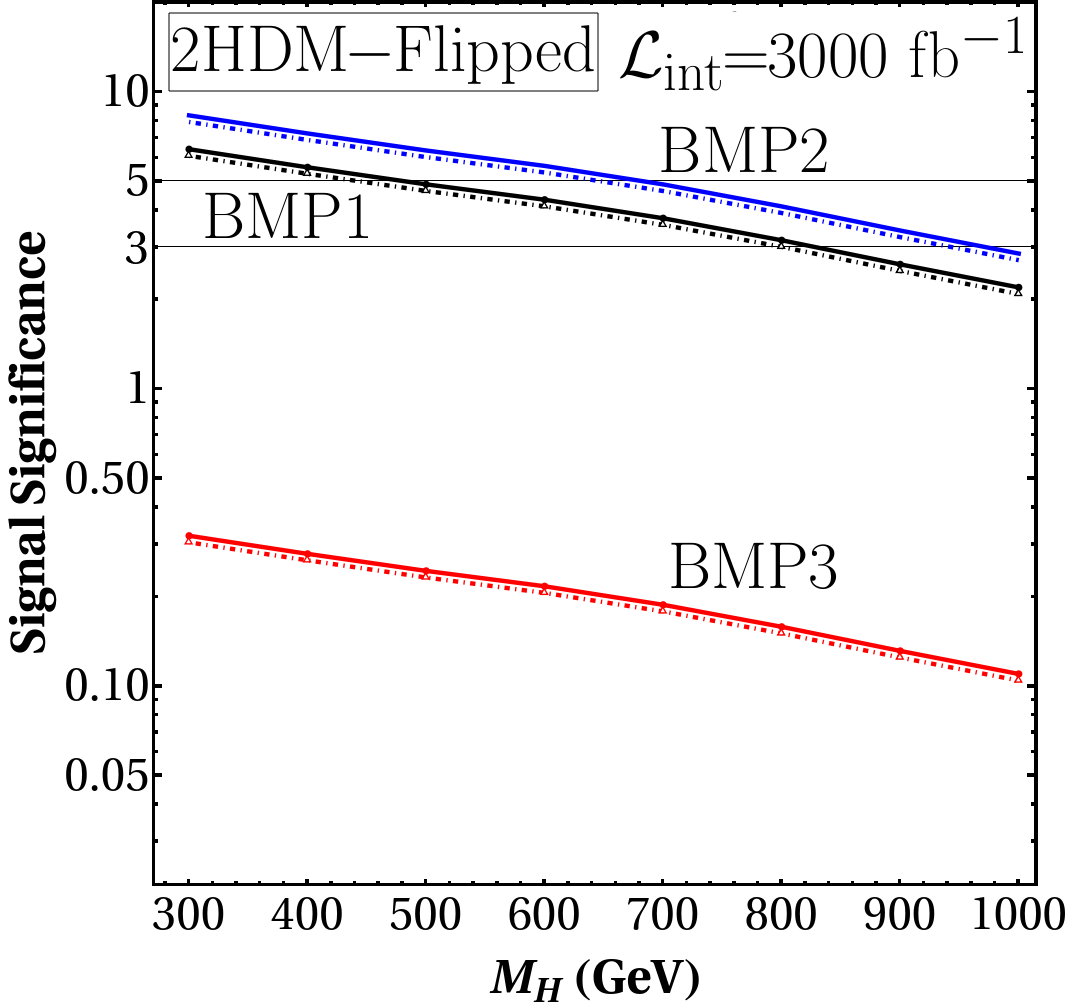}}
\end{center}
\caption{Signal significance expected at the LHC for the 2HDM-Flipped with an integrated luminosity of (a) 300, (b) 1000 and (c) 3000 fb$^{-1}$ for the process $H\to h \gamma\gamma$ (followed by $h\to b\bar{b})$, for the three benchmark points defined in the main text. Solid lines stand for the signal significance without systematic uncertainties, while the dotdashed lines consider a systematic uncertainty of $\kappa=5\%$.}
\label{Significance_2HDM-type_Flipped}
\end{figure}

\section{Conclusions}\label{Sec:conclusions}
We have presented a detailed study of the decay $H\to h \gamma\gamma$ at one-loop level in the framework of the Two-Higgs Doublet Model with natural flavor conservation,
concretely for the type-I, type-II, lepton specific and flipped 2HDMs. We shown that the LHC  could have sufficient sensitivity to detect signals of this rare $H$ decay, through the observation of the final state $\gamma\gamma b\bar b$ generated by the resonant production of the heavy CP even Higgs boson and the subsequent decay $h\rightarrow b\bar b$. 

 Specifically, we have identified regions of the model parameter space that predict a signal significance of $5\sigma$ for masses $M_H$ as high as $950$ GeV for the 2HDM-II. According to the analysis performed, the 2HDM-LS and 2HDM-Flipped would be the most difficult versions to distinguish through the $H\to h(b\bar{b})\gamma\gamma$ channel since the sensitivity differs by only a $5\%$ between then. In both versions, a significance of $5\sigma$ is predicted for $M_H\approx 650$ GeV. The least favored scenario corresponds to the 2HDM-I, which predicts detectable signals for masses up to 350 GeV. These results are obtained assuming an integrated luminosity of $3000$ fb$^{-1}$ and the BMP2 (the most favorable) corresponding to $\tan\beta=0.2$ for 2HDM-I, LS, flipped and $\tan\beta=5$ for 2HDM-II. BMP2 considers $\cos(\beta-\alpha)=0.01$, which represents a good approximation to the decoupling limit. In all cases we have considered a systematic uncertainty of 5$\%$.  
 
We also emphasize that virtual contributions are necessary to understand the $H\to b\bar{b}\gamma\gamma$ processes in the context of 2HDMs searches, since they generate a significant number of additional events, with non-trivial distribution, with respect to the two-body decay $H\to hh\,(h\to\gamma\gamma,\,h\to b\bar{b})$, thereby giving sensitivity to the rare decay presented in this work.

\subsection*{Acknowledgments}
The work of Marco A. Arroyo-Ure\~na and T. Valencia-P\'erez is supported by ``Estancias posdoctorales por M\'exico (SECIHTI)" and ``Sistema Nacional de Investigadores e Investigadoras" (SNII-SECIHTI). A.~Ibarra is supported by the Collaborative Research Center SFB1258 and by the Deutsche
Forschungsgemeinschaft (DFG, German Research Foundation) under Germany’s Excellence Strategy - EXC-2094 - 390783311. P.~Roig acknowledges Mexican support from CONAHCYT, particularly from project CBF2023-2024-3226, as well as Spanish support during his sabbatical through projects MCIN/AEI/10.13039/501100011033, grants PID2020-114473GB-
I00 and PID2023-146220NB-I00, and Generalitat Valenciana grant PROMETEO/2021/071. T.~V.~P. acknowledges support from the UNAM project PAPIIT IN111224 and the CONAHCYT project CBF2023-2024-548.

\subsection*{Note added}While we were preparing this manuscript, ref.~\cite{Phan:2024jbx} appeared, also studying the decay $H\to h\gamma\gamma$ in the Two Higgs Doublet Model.

	\appendix\label{Appendix}
	
\section{Feynman rules}\label{Acoplos}
In this appendix we present the Feynman rules involved in the calculations, which were performed in the unitary gauge. The direction of the momentum in the Feynman diagrams is indicated by arrows.
\begin{center}
\begin{table}
\includegraphics[height=0.15\textheight]{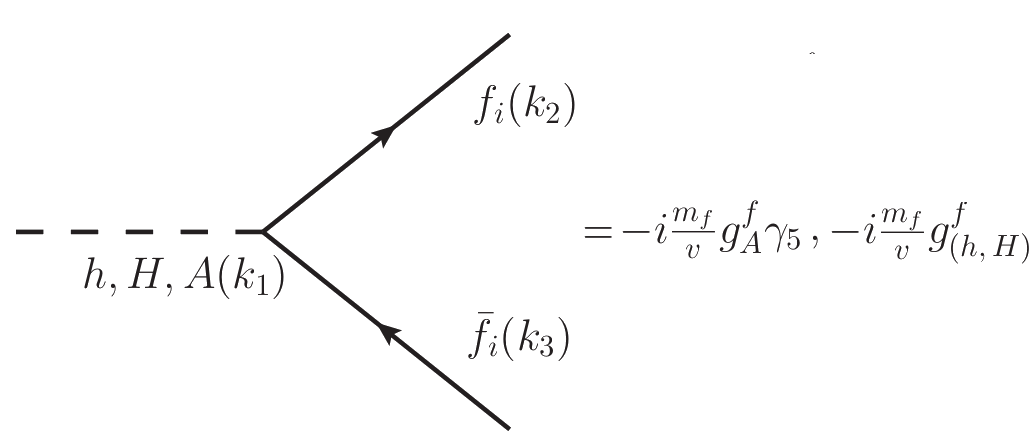}\\
\includegraphics[height=0.15\textheight]{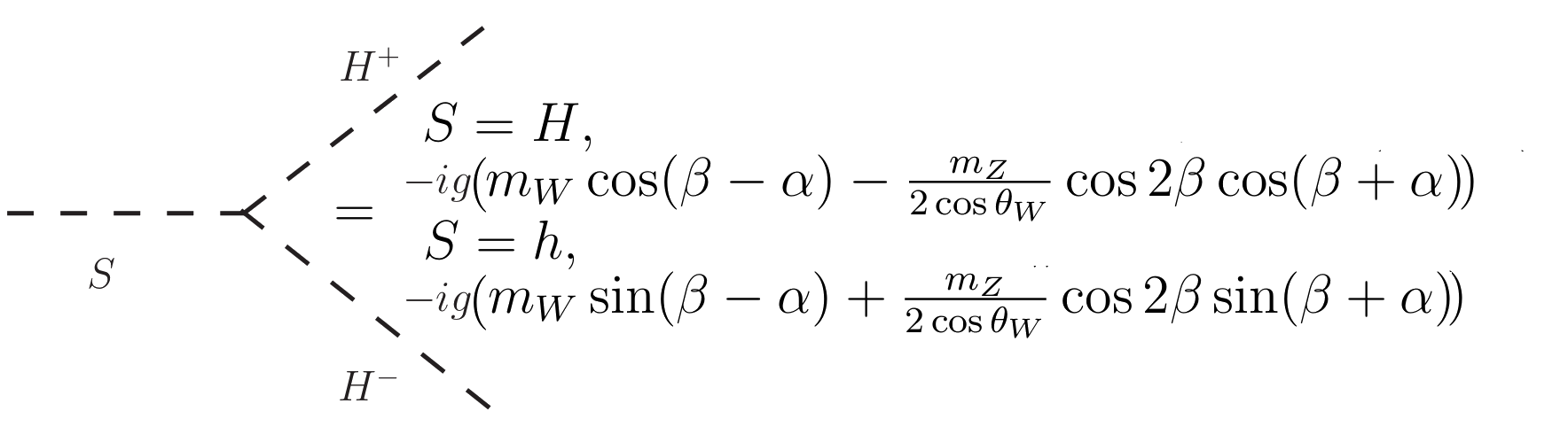}\\
\includegraphics[height=0.15\textheight]{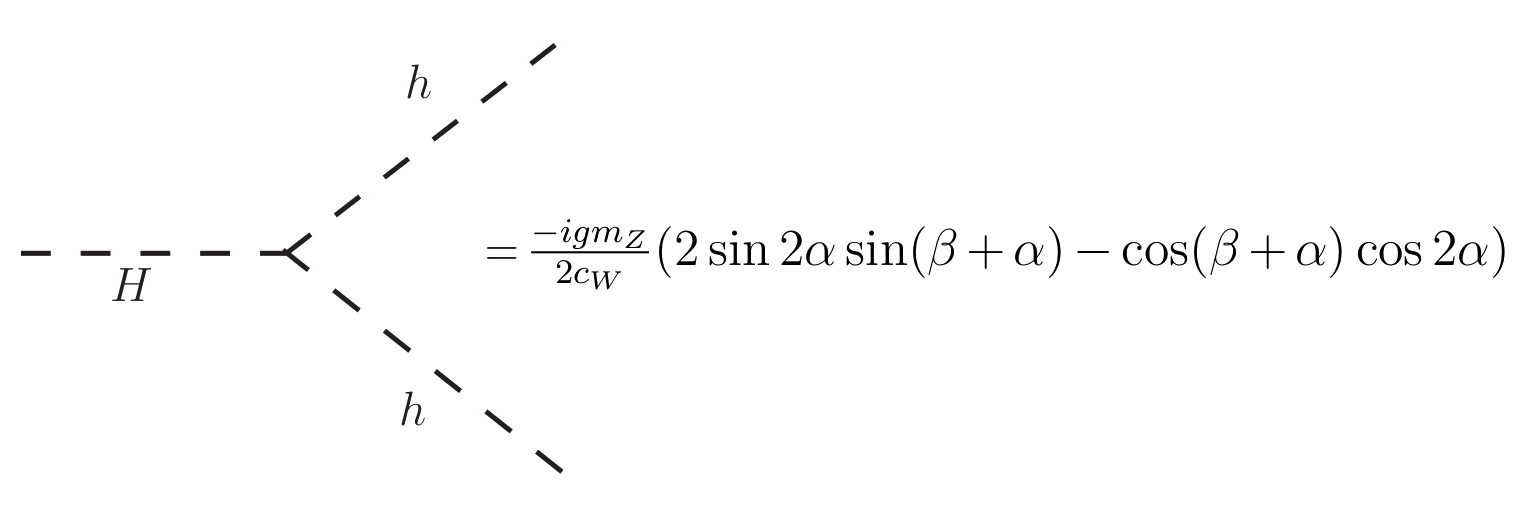}\\
\end{table}
\end{center}
\begin{center}
\begin{table}
\includegraphics[height=0.15\textheight]{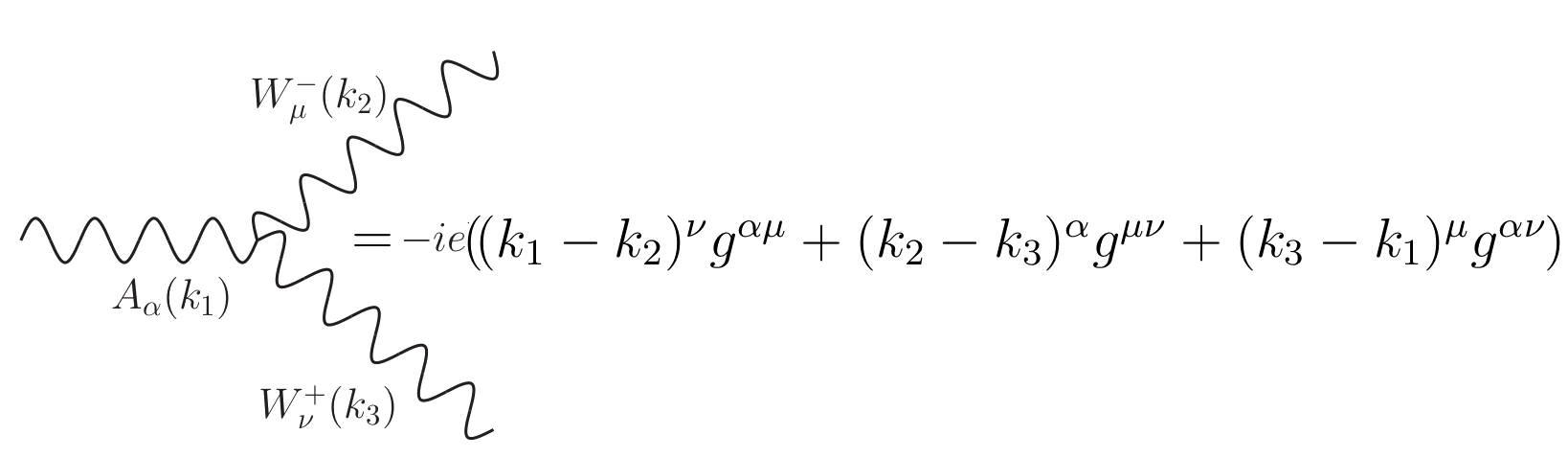}\\
\includegraphics[height=0.15\textheight]{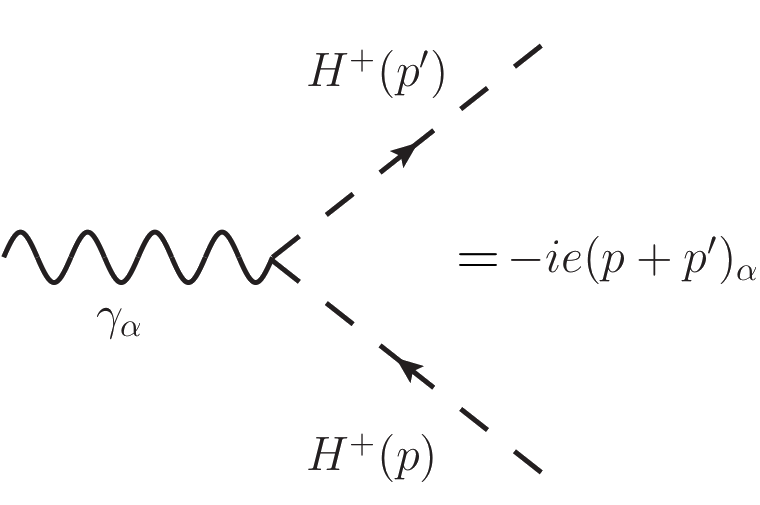}\\
\includegraphics[height=0.15\textheight]{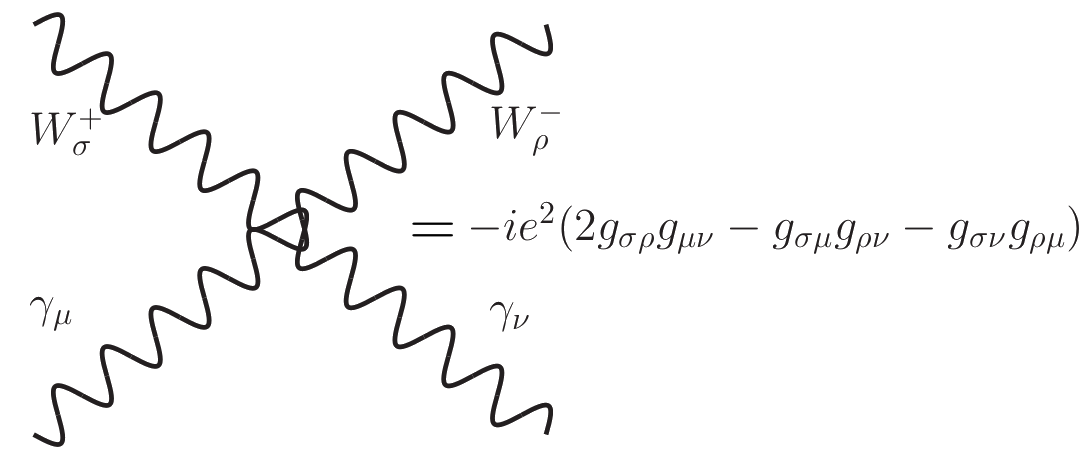}\\
\includegraphics[height=0.15\textheight]{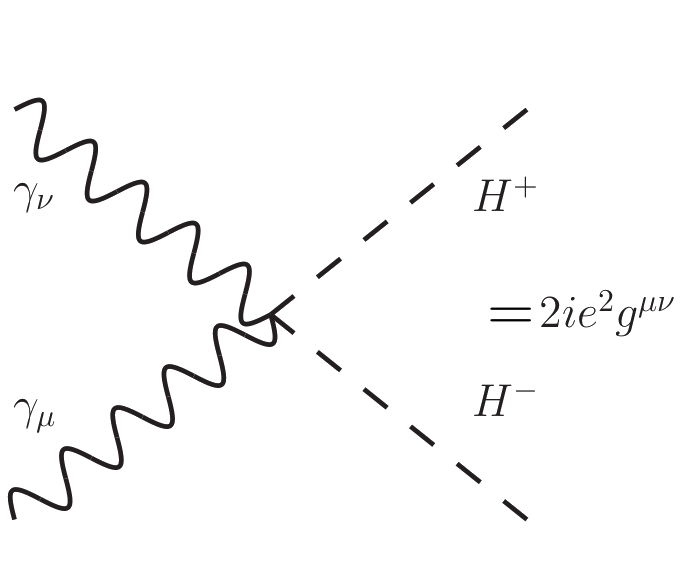}
\end{table}    
\end{center}

\section{Oblique parameters}\label{ObParam}
The $S,\,T,\,U$ oblique parameters in the theoretical framework of 2HDMs are given by
\begin{align}
		S  &=  \frac{1}{\pi m_{Z}^{2}}\Big\{ \sin^{2}\left(\beta-\alpha\right)\left[\mathcal{B}_{22}\left(m_{Z}^{2};m_{Z}^{2},m_{h}^{2}\right)-M_{Z}^{2}\mathcal{B}_{0}\left(m_{Z}^{2};m_{Z}^{2},m_{h}^{2}\right)+\mathcal{B}_{22}\left(m_{Z}^{2};M_{H}^{2},M_{A}^{2}\right)\right] \nonumber\\
		 &+  \cos^{2}\left(\beta-\alpha\right)\left[\mathcal{B}_{22}\left(m_{Z}^{2};m_{Z}^{2},M_{H}^{2}\right)-m_{Z}^{2}\mathcal{B}_{0}\left(m_{Z}^{2};m_{Z}^{2},M_{H}^{2}\right)+\mathcal{B}_{22}\left(m_{Z}^{2};m_{h}^{2},M_{A}^{2}\right)\right] \nonumber\\
		 &- \mathcal{B}_{22}\left(m_{Z}^{2};M_{H^{\pm}}^{2},M_{H^{\pm}}^{2}\right)-\mathcal{B}_{22}\left(m_{Z}^{2};m_{Z}^{2},m_{h}^{2}\right)+m_{Z}^{2}\mathcal{B}_{0}\left(m_{Z}^{2};m_{Z}^{2},m_{h}^{2}\right)\Big\} ,\\
		T & =  \frac{1}{16\pi m_{W}^{2}s_{W}^{2}}\Big\{ \sin^{2}\left(\beta-\alpha\right)\left[\mathcal{F}\left(M_{H^{\pm}}^{2},M_{H}^{2}\right)-\mathcal{F}\left(M_{H}^{2},M_{A}^{2}\right)+3\mathcal{F}\left(m_{Z}^{2},m_{h}^{2}\right)-3\mathcal{F}\left(m_{W}^{2},m_{h}^{2}\right)\right]\nonumber\\
		  &+ \cos^{2}\left(\beta-\alpha\right)\left[\mathcal{F}\left(M_{H^{\pm}}^{2},m_{h}^{2}\right)-\mathcal{F}\left(m_{h}^{2},M_{A}^{2}\right)+3\mathcal{F}\left(m_{Z}^{2},M_{H}^{2}\right)-3\mathcal{F}\left(m_{W}^{2},M_{H}^{2}\right)\right]\nonumber\\
		  &+\mathcal{F}\left(M_{H^{\pm}}^{2},M_{A}^{2}\right)-3\mathcal{F}\left(m_{Z}^{2},m_{h}^{2}\right)+3\mathcal{F}\left(m_{W}^{2},m_{h}^{2}\right)\Big\} ,\\
		U & =  \mathcal{H}\left(m_{W}^{2}\right)-\mathcal{H}\left(m_{Z}^{2}\right)\nonumber\\
		 &+  \frac{1}{\pi m_{W}^{2}}\Big\{ \cos^{2}\left(\beta-\alpha\right)\mathcal{B}_{22}\left(m_{W}^{2};M_{H^{\pm}}^{2},m_{h}^{2}\right)+\sin^{2}\left(\beta-\alpha\right)\mathcal{B}_{22}\left(m_{W}^{2};M_{H^{\pm}}^{2},M_{H}^{2}\right)\nonumber\\
		&+  \mathcal{B}_{22}\left(m_{W}^{2};M_{H^{\pm}}^{2},M_{A}^{2}\right)-2\mathcal{B}_{22}\left(m_{W}^{2};M_{H^{\pm}}^{2},M_{H^{\pm}}^{2}\right)\Big\} \nonumber\\
		 &- \frac{1}{\pi m_{Z}^{2}}\Big\{ \cos^{2}\left(\beta-\alpha\right)\mathcal{B}_{22}\left(m_{Z}^{2};m_{h}^{2},M_{A}^{2}\right)+\sin^{2}\left(\beta-\alpha\right)\mathcal{B}_{22}\left(m_{Z}^{2};M_{H}^{2},M_{A}^{2}\right)\nonumber\\
		 &-  \mathcal{B}_{22}\left(m_{Z}^{2};M_{H^{\pm}}^{2},M_{H}^{2}\right)\Big\} ,
\end{align}	
where
\begin{align}
		\mathcal{H}\left(m_{V}^{2}\right) & \equiv  \frac{1}{\pi m_{V}^{2}}\Big\{ \sin^{2}\left(\beta-\alpha\right)\left[\mathcal{B}_{22}\left(m_{V}^{2};m_{V}^{2},m_{h}^{2}\right)-m_{V}^{2}\mathcal{B}_{0}\left(m_{V}^{2};m_{V}^{2},m_{h}^{2}\right)\right]\nonumber\\
		& +  \cos^{2}\left(\beta-\alpha\right)\left[\mathcal{B}_{22}\left(m_{V}^{2};m_{V}^{2},M_{H}^{2}\right)-m_{V}^{2}\mathcal{B}_{0}\left(m_{V}^{2};m_{V}^{2},M_{H}^{2}\right)\right]\nonumber\\
		& -  \mathcal{B}_{22}\left(m_{V}^{2};m_{V}^{2},m_{h}^{2}\right)+m_{V}^{2}\mathcal{B}_{0}\left(m_{V}^{2};m_{V}^{2},m_{h}^{2}\right)\Big\} ,\nonumber\\
	\mathcal{F}\left(m_{1}^{2},m_{2}^{2}\right)&=\frac{1}{2}\left(m_{1}^{2}+m_{2}^{2}\right)-\frac{m_{1}^{2}m_{2}^{2}}{m_{1}^{2}-m_{2}^{2}}\log\left(\frac{m_{1}^{2}}{m_{2}^{2}}\right),\nonumber\\
	\mathcal{B}_{22}\left(q^{2};m_{1}^{2},m_{2}^{2}\right)&\equiv B_{22}\left(q^{2};m_{1}^{2},m_{2}^{2}\right)-B_{22}\left(0;m_{1}^{2},m_{2}^{2}\right),\nonumber\\
	\mathcal{B}_{0}\left(q^{2};m_{1}^{2},m_{2}^{2}\right)&\equiv B_{0}\left(q^{2};m_{1}^{2},m_{2}^{2}\right)-B_{0}\left(0;m_{1}^{2},m_{2}^{2}\right),\nonumber\\
		B_{22}\left(q^{2};m_{1}^{2},m_{2}^{2}\right) & =  \frac{1}{6}\Big[A_{0}\left(m_{2}^{2}\right)+2m_{1}^{2}B_{0}\left(q^{2};m_{1}^{2},m_{2}^{2}\right)+\left(m_{1}^{2}-m_{2}^{2}+q^{2}\right)B_{1}\left(q^{2};m_{1}^{2},m_{2}^{2}\right)\nonumber\\
		& - \frac{q^{2}}{3}+m_{1}^{2}+m_{2}^{2}\Big],
\end{align}
and $A_0,\,B_0,\,B_1$ are scalar Passarino-Veltman functions.


\section{Amplitude}\label{amplitudes}
Here we present the exact one-loop expression for the amplitude of the process $H(p)\to \gamma(k_1) \gamma(k_2) h(k_3)$, summing over the polarizations of the final photons. The result is:
\begin{align}\label{M2phiDecay}
	\overline{|\mathcal{M}(H\to h\gamma\gamma)|^2}&=
	\chi_2(\left| G_1\right| ^2
	-{\rm Re}[G_1\tilde{G}_1^*])
	- \chi_1(\hat\Theta_{1h}{\rm Re}\left[G_1 G_2^*\right]
	+ \hat\Theta _{2 h}{\rm Re}[G_1 \tilde{G}_2^*])
		\nonumber\\
	&+\mu_3\left(\hat\Theta_{1h}{\rm Re}\left[G_1 G_3^*\right]
	-\hat\Theta _{2 h}{\rm Re}[G_1 \tilde{G}_3^*]\right)
	+\chi_2({\rm Re}[G_1 G_4^*]
	- {\rm Re}[G_1 \tilde{G}_4^*])
	\nn\\
	&+\chi_1\Big(-\frac{1}{4} \hat\Theta_{1h}^2 \left| G_2\right| ^2
	+\frac{1}{8} \hat\Theta_{1h} \hat\Theta _{2 h} {\rm Re}[G_2\tilde{G}_2^*]
	+\frac{1}{2} \mu_3{\rm Re}[ G_2 \tilde{G}_3^*]
	-\hat\Theta_{1h}{\rm Re}[G_2 G_4^*]\Big)
		\nonumber\\
	&+\mu_3\Big(\hat\Theta_{1h}\chi_1\frac{1}{2} {\rm Re}[G_2 \tilde{G}_4^*]
	+\frac{1 \hat\Theta_{1h}^2}{4 \mu _h} \left| G_3\right| ^2\Big)
	+\frac{1}{2 \mu _h}  [\hat\Theta_{1h} \hat\Theta _{2 h}-2 \mu_3 \mu _h]{\rm Re}[ G_3 \tilde{G}_3^*]
	\nonumber\\
	&+\frac{1}{2} \mu_3  \hat\Theta_{1h} {\rm Re}\left[G_3 G_4^*\right]
	-\hat\Theta_{1h}{\rm Re}[G_3 \tilde{G}_4^*]
	+\frac{\chi_3}{4\mu_3 \mu _h}(\left| G_4\right| ^2
	-{\rm Re}[G_4 \tilde{G}_4^*])
	\nn\\
	&+\left(\mu_1\leftrightarrow\mu_2\right)
	,
\end{align}

where $\mu_h=m_h^2/M_H^2$ and $\mu_i=L_i/M_H^2$, with $L_i$ the Lorentz invariant quantities
\begin{align}
L_1&=(k_1+k_3)^2,\nonumber\\	
L_2&=(k_2+k_3)^2,\nonumber\\
L_3&=(k_1+k_2)^2,
\end{align} 
which satisfy $L_1+L_2+L_3=M_H^2+M_h^2$. Further, the other quantities in Eq.~(\ref{M2phiDecay}) are:
\begin{align}
  G_i&=G_i(\mu_3,\mu_2,\mu_1),\nn \\
  \tilde{G}_i&=G_i(\mu_3,\mu_1,\mu_2),\nn \\
  \Theta_{ij}&=L_i-m_j^2,\nn \\
   \hat{\Theta}_{ij}&=\Theta_{ij}/M_H^2
 ,\nn\\
	\chi_1&=\mu_3\mu_h-\hat\Theta_{1h}\hat\Theta_{2h},\nn\\
	\chi_2&=2\hat\Theta_{1h}\hat\Theta_{2h}-\mu_3\mu_h,\nn\\
	\chi_3&=\mu_1^2\hat\Theta_{2h}^2-2\mu_1(\hat\Theta_{2h}-\mu_3)\mu_h\hat\Theta_{2h},
\end{align}
where
\begin{align}
	G_1&= \sum_f\alpha_f \mathcal{B}_f^{\rm box} ,\nn\\
	G_2&= \alpha_W \mathcal{B}_W^{\rm box}, \nn \\
    G_3&= \alpha_{H^\pm} \mathcal{B}_{H^\pm}^{\rm box}, \nn \\
    G_4&= \mathcal{B}^{\rm rd},
\end{align}
are respectively the form factors induced by  box diagram with charged fermions, $W$ bosons and charged Higgses, and the form factor induced by the reducible diagrams. Here,
\begin{align}
	\alpha_f&=\frac{m_f^2 g_h^f g_H^f e^2Q_f^2}{v^2},\nn\\
	\alpha_W&=\frac{e^4m_W^2 g_h^{WW} g_H^{WW}}{s_W^2},\nn\\
	\alpha_{H^\pm} &=\frac{e^4m_W^2 g_h^{H^-H^+} g_H^{H^-H^+}}{s_W^2},
\end{align}
and
\begin{align}
		\mathcal{B}_f^{\rm box} &=\frac{2}{L_3}\Big\{-\Big((L_1L_2-m_h^2m_f^2) +L_3(L_1+L_2)\Big)\textbf{C}_1^f\nn \\ &+\Big(\Theta_{1h}\Theta_{1H} +L_2\Theta_{2H}\Big)\textbf{C}_2^f\nonumber\\
	&-\Big(2\Theta_{1h}L_3+\dfrac{2}{L_3}\Theta_{2h}^2\Theta_{2H 
        }\Big)\textbf{C}_3^f\nn \\
        &+\Big(4(L_1L_2-m_h^2m_f^2)+2(L_1-L_2)^2\Big)\textbf{C}_4^f\nonumber\\
		& +\Big(L_1((L_1L_2-m_h^2m_f^2) +2L_1L_3)+L_2((L_1L_2-m_h^2m_f^2) +2L_2L_3)\Big)\textbf{D}_1^f \nonumber\\
		&+\dfrac{1}{L_3}\Big( (L_1L_2-m_h^2m_f^2) ((L_1L_2-m_h^2m_f^2) +4L_3m_f^2)\Big)\textbf{D}_2^f\Big\},\\ \nn  \\
		\mathcal{B}_W^{\rm box} &=\dfrac{1}{L_3^2m_W^4}\Big\{-L_3\Big(L_1L_2+M_H ^2(m_h^2+2L_2)+\Theta_{2h}L_2 \Big)\textbf{C}_1^W\nn \\
  &-\Theta_{1H}\Big( m_h^4-(3 L_3+2(L_1+L_2)) m_h^2
		+(2 L_3^2+3 L_1 L_3+L_1^2+L_2^2+4 (L_3+L_1) L_2)\nn \\
		&-(L_1+L_2) L_3+ (4L_1+L_2)L_2+2 ((L_1+L_2)+(L_3+L_1))L_1   +\Theta_{2H}( m_h^2-L_2) \Big)\textbf{C}_2^W \nonumber\\
		&+\Big[\Theta_{2h}\Big((L_1L_2-m_h^2m_W^2) -L_3m_h^2\Big)  +\Big(L_3m_h^2(L_3+2L_2-m_h^2) -\Theta_{1h}\Theta_{2h}^2 \Big)\Big]\textbf{C}_3^W \nonumber\\
		&-m_h^2\Big( 2(L_1L_2-m_h^2m_W^2) +(L_1-L_2)^2  \Big)\textbf{C}_4^W\nn \\
  &+\frac{1}{2} \Big[4 (L_1L_2-m_h^2m_W^2) (L_1L_2-m_h^2m_W^2)-L_3m_h^4 m_W^2)+L_3 \Big(-2
		m_h^6+m_h^4(4 L_3+3 L_1+4 L_2)\nn \\
  &-m_h^2(2 L_3^2+3
		L_1 L_3+4 L_2 L_3+L_1^2+2 L_2^2
		+6 L_1 L_2) +m_h^2
		(3 L_2^2+3 L_3 L_2+2 L_1 L_2-2 L_3 L_1)\nn \\
		&-L_1 L_2 L_3\Big)+
		\frac{1}{2} L_3^2 \Big(4 (L_1L_2-m_h^2m_W^2)+ (L_1L_2-m_h^2m_W^2) -L_3 m_W^2\Big)+L_3 L_2\Big(m_h^2(L_2+M_H^2)\nn \\
  &-L_2 (3
		L_3+2 L_1+L_2) +L_1 L_2\Big)\Big]\textbf{D}_1^W \nonumber\\
		&+\frac{1}{2} (L_1L_2-m_h^2m_W^2) \Big(2 m_h^4-(5 L_1 L_3+6 L_2 L_3+2 L_1^2
		+2 L_2^2+8 L_1 L_2)
		-4  (2 L_3+L_1+L_2) m_W^2\nn \\&+
		(L_3^2+(6 L_1+L_2) L_3+4 L_1
		(L_1+L_2))+ 4
		L_3 m_W^2+(L_3+2 L_1) L_2\Big) \textbf{D}_2^W
		\Big\},\\ \nn  \\
		\mathcal{B}_{H^\pm}^{\rm box}&=\frac{1}{16(L_1L_2-m_h^2m_{H^{\pm}}^2)}\Big[\Theta_{2H}\textbf{C}_1^{H^{\pm}}-\Theta_{1H}\textbf{C}_2^{H^{\pm}}+2(L_1-L_2)\textbf{C}_4^{H^{\pm}}\nn \\
  &+(\Theta_{2H}\Theta_{2h}-\Theta_{1H}\Theta_{1h})\textbf{D}_1^{H^{\pm}}\Big],\\ \nn  \\
        \mathcal{B}^{ \rm \LARGE rd} &=\dfrac{1}{8\Theta_{1h}\Theta_{2h}\Theta_{3h}}\Big\{ 4m_h^2(L_1L_2-m_h^2(m_f^2+m_W^2+m_{H^-}^2))[ \delta \textbf{B}(m_h^2,\tilde{L}_2)+\delta \textbf{B}(m_h^2,\tilde{L}_1)]  \nonumber\\
		&+4\Theta_{2h}(L_1L_2-m_h^2(m_f^2+m_W^2+m_{H^-}^2)) \delta \textbf{B}(\tilde{L}_3,L_3)\nn \\
  &+2 L_3\Theta_{1h}\Big[2(L_1L_2-m_h^2m_f^2)+m_h^2(m_h^2-L_3)-L_1(L_2
		+L_1) +L_2^2\Big]\textbf{C}_1^f\nn \\&+\Theta_{1h}
		(L_3+\Theta_{2h})\Big[2L_1\Theta_{1h}-(L_1L_2 -m_h^2m_f^2)\Big]\textbf{C}_2^f\nn \\
  &+\Theta_{1h}\Theta_{2h}
		\Big[5(L_1L_2-m_h^2m_f^2)-2L_2(\Theta_{1h}+L_1)+2L_2^2 \Big]\textbf{C}_3^f\nn \\
  &-\Theta_{2h}(L_3+\Theta_{1h})
		\Big[5(L_1L_2-m_h^2m_W^2)-2L_2(\Theta_{1h}+L_1)+2L_2^2 \Big]\textbf{C}_2^W \nonumber\\
		&-\Theta_{1h}\Theta_{2h}
		\Big[2L_1\Theta_{1h}-(L_1L_2-m_h^2m_W^2)\Big]\textbf{C}_3^W\nn \\
  &+2 \Theta_{1h}  \Theta_{2h}\Big[2m_h^2(M_H^2+2(L_1+L_2))-5 L_1^2-3	L_2^2+4 L_3 (L_1-L_2) +(L_1-L_2)(L_1+L_2)\Big]\textbf{C}_4^{H^{\pm}}\nn \\
  &+\Theta_{1h}\Theta_{2h}\Big[-2m_h^4(m_h^2+2M_H^2)+m_h^2(2 L_3^2+(5	L_1+4 L_2) L_3 +2 (L_1^2+4 L_2 L_1+L_2^2))\nn \\
  &-L_1 (L_1^2-(L_1-5
		L_2) L_3+4 L_2 (L_1+L_2))+L_1 (2 L_2^2+L_3 (L_2-2L_1))-4 \Theta_{1h}	(L_1L_2-m_h^2m_f^2)\Big]\textbf{D}_1^f\nn \\
  &-\Theta_{1h}\Theta_{2h}\Big[-2m_h^4(m_h^2+2M_H^2)+m_h^2[2(L_3+L_1)^2+2 L_2^2+(9 L_3+8 L_1)	L_2)-L_3 (5 L_3+3 (3L_1+L_2))\nn \\
  & +4 L_1 (L_1+L_2)]+L_2 (L_1 (L_3+2 L_1)+2 L_3L_2)+4 m_W^2 m_h^2 (m_h^2-2 L_1+L_2)+m_h^2(L_1L_2-m_h^2m_W^2)\Big]\textbf{D}_1^W \nonumber\\
		&+4 m_{H^{\pm}}^2 \Theta_{1h}\Theta_{2h}
		(L_1L_2-m_h^2m_{H^-}^2) \textbf{D}_2^{H^{\pm}}
		\Big\},
 \end{align}
where the three- and four-point Passarino-Veltman scalar functions, $\textbf{C}_i^P$ and $\textbf{D}_i^P$ ($P=f,\,W,\,H^{\pm}$), are defined as
\begin{align}
	\textbf{C}_1^P &= C_0(0, 0, M_H^2, m_P^2, m_P^2, m_P^2), \nonumber\\
	\textbf{C}_2^P&= C_0(0,M_H^2, M_H^2 , m_P^2, m_P^2, m_P^2),\nonumber\\
	\textbf{C}_3^P&= C_0(m_h^2, 0,M_H^2, m_P^2, m_P^2, m_P^2),\nonumber\\
	\textbf{C}_4^P &= C_0(m_h^2, M_H^2, M_H^2 , m_P^2, m_P^2, m_P^2), \nonumber\\
	\textbf{D}_1^P &= D_0(m_h^2, 0, 0, M_H^2 , M_H^2, L_3,m_P^2, m_P^2, m_P^2, m_P^2),\nonumber\\
	\textbf{D}_2^P &= D_0(m_h^2, 0, M_H^2, 0,L_1, m_P^2, m_P^2, m_P^2, m_P^2, m_P^2).
\end{align}
Meanwhile, the two-point scalar functions are defined as $\delta \textbf{B}(x,y)=B_0(x,\,m_f^2,\,m_f^2)-B_0(y,\,m_f^2,\,m_f^2)$ and $\tilde{L}_1=L_1-2k_1\cdot k_3,
  \tilde{L}_2=L_2-2k_2\cdot k_3$, $\tilde{L}_3=2k_1\cdot k_2$. It is clear that ultraviolet divergences cancel out.

\bibliographystyle{apsrev4-1}
\bibliography{biblio}

\end{document}